\documentclass[twocolumn,floatfix,aps,superscriptaddress,eqsecnum]{revtex4}
\usepackage{graphicx}
\usepackage{amsmath}
\usepackage{amssymb}
\usepackage{bm}

\usepackage{color}

\begin{document}

\title{Magnetic breakdown spectrum of a Kramers-Weyl semimetal}
\author{G. Lemut}
\affiliation{Instituut-Lorentz, Universiteit Leiden, P.O. Box 9506, 2300 RA Leiden, The Netherlands}
\author{A. Don\'{i}s Vela}
\affiliation{Instituut-Lorentz, Universiteit Leiden, P.O. Box 9506, 2300 RA Leiden, The Netherlands}
\author{M. J. Pacholski}
\affiliation{Instituut-Lorentz, Universiteit Leiden, P.O. Box 9506, 2300 RA Leiden, The Netherlands}
\author{J. Tworzyd{\l}o}
\affiliation{Faculty of Physics, University of Warsaw, ul.\ Pasteura 5, 02--093 Warszawa, Poland}
\author{C. W. J. Beenakker}
\affiliation{Instituut-Lorentz, Universiteit Leiden, P.O. Box 9506, 2300 RA Leiden, The Netherlands}

\date{April 2020}
\begin{abstract}
We calculate the Landau levels of a Kramers-Weyl semimetal thin slab in a perpendicular magnetic field $B$. The coupling of Fermi arcs on opposite surfaces broadens the Landau levels with a band width that oscillates periodically in $1/B$. We interpret the spectrum in terms of a one-dimensional superlattice induced by magnetic breakdown at Weyl points. The band width oscillations may be observed as $1/B$-periodic magnetoconductance oscillations, at weaker fields and higher temperatures than the Shubnikov-de Haas oscillations due to Landau level quantization. No such spectrum appears in a generic Weyl semimetal, the Kramers degeneracy at time-reversally invariant momenta is essential.
\end{abstract}
\maketitle

\section{Introduction}
\label{intro}

Kramers-Weyl fermions are massless low-energy excitations that may appear in the Brillouin zone near time-reversally invariant momenta (TRIM). Their gapless nature is protected by Kramers degeneracy, which enforces a band crossing at the TRIM. Crystals that support Kramers-Weyl fermions have strong spin-orbit coupling and belong to one of the chiral point groups, without reflection or mirror symmetry, to allow for a linear rather than quadratic band splitting away from the TRIM. The materials are called topological chiral crystals or Kramers-Weyl semimetals --- to be distinguished from generic Weyl semimetals where Kramers degeneracy plays no role. Several candidates were predicted theoretically \cite{Cha18,She18} and some have been realized in the laboratory \cite{Rao19,San19,Tak19,Yua19,Sch19}.  

These recent developments have motivated the search for observables that would distinguish Kramers-Weyl fermions from generic Weyl fermions \cite{Zha17,Wa18,He19}. Here we report on the fundamentally different Landau level spectrum when the semimetal is confined to a thin slab in a perpendicular magnetic field.

Generically, Landau levels are dispersionless: The energy does not depend on the momentum in the plane perpendicular to the magnetic field $B$. In contrast, we have found that the Landau levels of a Kramers-Weyl semimetal are broadened into a Landau band. The band width oscillates periodically in $1/B$, producing an oscillatory contribution to the magnetoconductance. 

\begin{figure*}[tb]
\centerline{\includegraphics[width=0.7\linewidth]{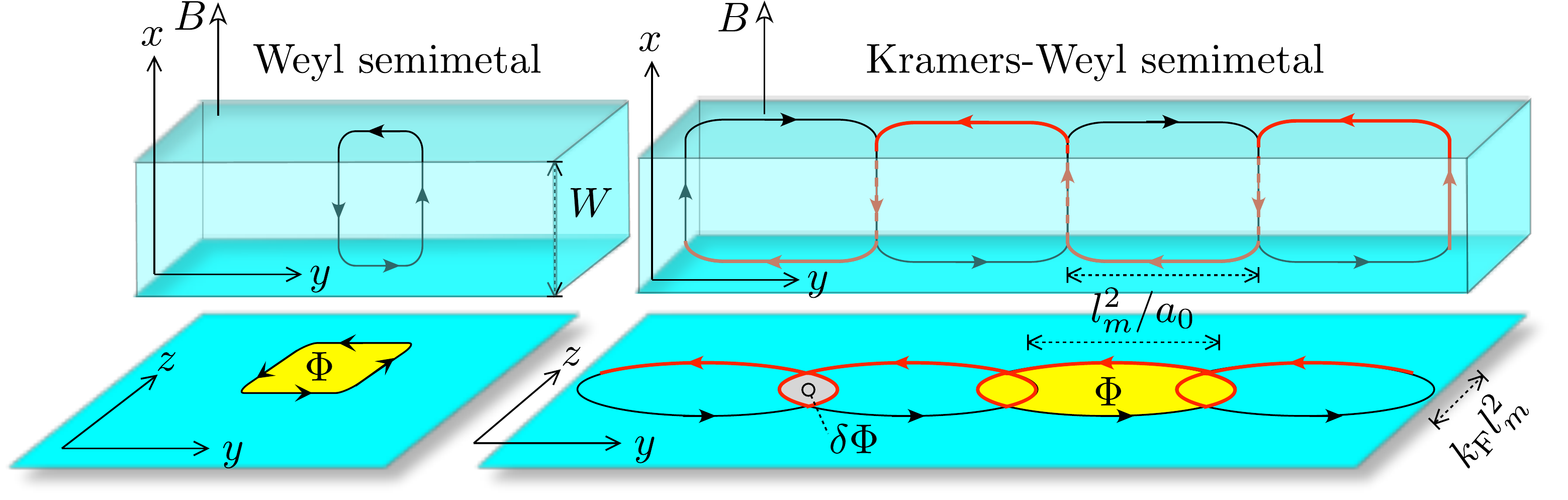}}
\caption{Electron orbits in a thin slab geometry perpendicular to a magnetic field (along the $x$-axis), for a generic Weyl semimetal \cite{Pot14,Zha16} (at the left) and for a Kramers-Weyl semimetal (at the right). In each case we show separately a front view (in the $x$--$y$ plane, to show how the orbits switch between top and bottom surfaces of the slab) and a top view (in the $y$--$z$ plane, to indicate the magnetic flux enclosed by the orbits). The Kramers-Weyl semimetal combines open orbits (red arrows) with closed orbits enclosing either a large flux $\Phi$ or a small flux $\delta\Phi$. Open and closed orbits are coupled by a periodic chain of magnetic breakdown events, spaced by $l_m^2/a_0$ (with $a_0$ the lattice constant and $l_m=\sqrt{\hbar/eB}$ the magnetic length). The open orbits broaden the Landau levels into a band, the band width varies from minimal to maximal when $\delta\Phi$ is incremented by $h/e$. Because $\delta\Phi\propto Bl_m^4\propto 1/B$, the band width oscillations are periodic in $1/B$.
}
\label{fig_conveyor}
\end{figure*}

The phenomenology is similar to that encountered in a semiconductor 2D electron gas in a superlattice potential \cite{Ger89,Win89,Bee89,Str90,Gvo07}. In that system the dispersion is due to the drift velocity of cyclotron orbits in perpendicular electric and magnetic fields. Here the surface Fermi arcs provide for open orbits, connected to closed orbits by magnetic breakdown at Weyl points (see Fig.\ \ref{fig_conveyor}). 

No open orbits appear in a generic Weyl semimetal \cite{Pot14,Zha16}, because the Weyl points are closely separated inside the first Brillouin zone, so the Fermi arcs are short and do not cross the Brillouin zone boundaries (a prerequisite for open orbits). The Landau band dispersion therefore directly ties into a defining property \cite{Cha18} of a Kramers-Weyl semimetal: surface Fermi arcs that span the entire Brillouin zone because they connect TRIM at zone boundaries.

In the next two sections \ref{sec_Hbc} and \ref{sec_Fermisurface} we first compute the spectrum of a Kramers-Weyl semimetal slab in zero magnetic field, to obtain the equi-energy contours that govern the orbits when we apply a perpendicular field. The resonant tunneling between open and closed orbits via magnetic breakdown is studied in Sec.\ \ref{sec_openclosed}. With these preparations we are ready to calculate the dispersive Landau bands and the magnetoconductance oscillations in Secs.\ \ref{sec_dispersiveLL} and \ref{sec_magneto}. The analytical calculations are then compared with the numerical solution of a tight-binding model in Secs.\ \ref{sec-tbmodel} and \ref{sec_numerics}. We conclude in Sec.\ \ref{sec_conclude}.

\section{Boundary condition for Kramers-Weyl fermions}
\label{sec_Hbc}

The first step in our analysis is to characterize the surface Fermi arcs in a Kramers-Weyl semimetal, which requires a determination of the boundary condition for Kramers-Weyl fermions. This is more strongly constrained by time-reversal symmetry than the familiar boundary condition on the Dirac equation \cite{Akh08}. In that case the confinement by a Dirac mass $V_\mu=\mu(\hat{\bm n}_{\parallel}\cdot\bm{\sigma})$ generates a boundary condition
\begin{equation}
\Psi=(\hat{\bm n}_\perp\times\hat{\bm n}_{\parallel})\cdot\bm{\sigma}\Psi.\label{Psisinglecone}
\end{equation}
The unit vectors $\hat{\bm n}_\parallel$ and $\hat{\bm n}_\perp$ are parallel and perpendicular to the boundary, respectively.

Although $\bm{\sigma}\mapsto-\bm{\sigma}$ upon time reversal, the Dirac mass may still preserve time-reversal symmetry if the Weyl fermions are not at a time-reversally invariant momentum (TRIM). For example, in graphene a Dirac mass $+\mu$ at the K-point in the Brillouin zone and a Dirac mass $-\mu$ at the $\text{K}'$-point preserves time-reversal symmetry. In contrast, for Kramers-Weyl fermions at a TRIM the $V_\mu$ term in the Hamiltonian is incompatible with time-reversal symmetry. To preserve time-reversal symmetry the boundary condition must couple two Weyl cones, it cannot be of the single-cone form \eqref{Psisinglecone}.

In App.\ \ref{sec_proof} we demonstrate that, indeed, pairs of Weyl cones at a TRIM are coupled at the boundary of a Kramers-Weyl semimetal. Relying on that result, we derive in this section the time-reversal invariant boundary condition for Kramers-Weyl fermions.

We consider a Kramers-Weyl semimetal in a slab geometry, confined to the $y$--$z$ plane by boundaries at $x=0$ and $x=W$. In a minimal description we account for the coupling of two Weyl cones at the boundary. To first order in momentum $\bm{k}$, measured from a Weyl point, the Hamiltonian of the uncoupled Weyl cones is
\begin{equation}
\begin{split}
&H_\pm(\bm{k})=\begin{pmatrix}
H_0(\bm{k})+\varepsilon&0\\
0&\pm H_0(\bm{k})-\varepsilon
\end{pmatrix},\\
&H_0(\bm{k})=\textstyle{\sum_{\alpha=x,y,z}}v_\alpha k_\alpha\sigma_\alpha.
\end{split}
\end{equation}
The $\pm$ sign indicates whether the two Weyl cones have the same chirality ($+$) or the opposite chirality ($-$). The two Weyl points need not be at the same energy, we allow for an offset $\varepsilon$. We also allow for anisotropy in the velocity components $v_\alpha$. 

The $\sigma_\alpha$'s are Pauli matrices acting on the spin degree of freedom. We will also use $\tau_\alpha$ Pauli matrices that act on the Weyl cone index, with $\sigma_0$ and $\tau_0$ the corresponding $2\times 2$ unit matrix. We can then write
\begin{equation}
H_+=H_0\tau_0+\varepsilon\tau_z,\;\;H_-=H_0\tau_z+\varepsilon\tau_z.
\end{equation}
The current operator in the $x$-direction is $j_+=v_x\sigma_x\tau_0$ for $H_+$ and $j_-=v_x\sigma_x\tau_z$ for $H_-$. The time-reversal operation ${\cal T}$ does not couple Weyl cones at a TRIM, it only inverts the spin and momentum:
\begin{equation}
{\cal T}H_\pm(\bm{k}){\cal T}^{-1}=\sigma_y H^\ast_\pm(-\bm{k})\sigma_y=H_\pm(\bm{k}).
\end{equation}

An energy-independent boundary condition on the wave function $\Psi$ has the general form \cite{Akh08}
\begin{equation}
\Psi=M_\pm\cdot\Psi,\;\;M_\pm=M_\pm^\dagger,\;\;M_\pm^2=1,
\end{equation}
in terms of a Hermitian and unitary matrix $M_\pm$. The matrix $M_\pm$ anticommutes with the current operator $j_\pm$ perpendicular to the surface, to ensure current conservation. Time-reversal symmetry further requires that
\begin{equation}
\sigma_y M_\pm^\ast\sigma_y=M_\pm.
\end{equation}
These restrictions reduce $M_\pm$ to the single-parameter form
\begin{equation}
\begin{split}
&M_+(\phi)=\tau_y\sigma_y\cos\phi+\tau_y\sigma_z\sin\phi,\\
&M_-(\phi)=\tau_x\sigma_0\cos\phi +\tau_y\sigma_x\sin\phi.
\end{split}\label{Mpmdef}
\end{equation}

The angle $\phi$ has a simple physical interpretation in the case $H_+,M_+$ case of two coupled Weyl cones of the same chirality: It determines the direction of propagation of the helical surface states (the Fermi arcs). We will take $\phi=0$ at $x=0$ and $\phi=\pi$ at $x=W$. This produces a surface state that is an eigenstate of $\tau_y\sigma_y$ with eigenvalue $+1$ on one surface and eigenvalue $-1$ on the opposite surface, so a circulating surface state in the $\pm y$-direction. (Alternatively, if we would take $\phi=\pm\pi/2$ the state would circulate in the $\pm z$-direction.)

Notice that these are helical rather than chiral surface states: The eigenstates $\Psi$ of $\tau_y\sigma_y$ with eigenvalue $+1$ contain both right-movers ($\sigma_y\Psi=+\Psi$) and left-movers ($\sigma_y\Psi=-\Psi$). This is the key distinction with surface states in a magnetic Weyl semimetal, which circulate unidirectionally around the slab \cite{Has17,Yan17,Bur18,Arm18}.

In the case $H_-,M_-$ that the coupled Weyl cones have the opposite chirality there are no helical surface states and the physical interpretation of the angle $\phi$ in Eq.\ \eqref{Mpmdef} is less obvious. Since our interest here is in the Fermi arcs, we will not consider that case further in what follows.

\section{Fermi surface of Kramers-Weyl fermions in a slab}
\label{sec_Fermisurface}

\subsection{Dispersion relation}
\label{sec_dispersion}

We calculate the energy spectrum of $H_+$ with boundary condition $M_+$ from Eq.\ \eqref{Mpmdef} along the lines of Ref.\ \onlinecite{Bov18}. Integration in the $x$-direction of the wave equation $H_\pm\Psi=E\Psi$ with $k_x=-i\hbar \partial/\partial x$ relates the wave amplitudes at the top and bottom surface via $\Psi(W)=e^{i\Xi}\Psi(0)$, with
\begin{equation}
\Xi=\frac{W}{\hbar v_x}\sigma_x(E-v_yk_y\sigma_y-v_zk_z\sigma_z-\varepsilon\tau_z).
\end{equation}
As discussed in Sec.\ \ref{sec_Hbc} we impose the boundary condition $\Psi=M_+(0)\Psi$ on the $x=0$ surface and $\Psi=M_+(\pi)\Psi$ on the $x=W$ surface.

The round-trip evolution
\begin{equation}
\Psi(0)=M_+(0)e^{-i\Xi}M_+(\pi)e^{i\Xi}\Psi(0)
\end{equation}
then gives the determinantal equation
\begin{equation}
{\rm Det}\,\left(1+\tau_y\sigma_y e^{-i\Xi}\tau_y\sigma_y e^{i\Xi}\right)=0,
\end{equation}
which evaluates to
\begin{align}
[E^2-\varepsilon^2+(v_z k_z)^2-(v_y k_y)^2]\frac{\sin w_- \sin w_+}{q_-q_+}\nonumber\\
=1+\cos w_- \cos w_+,\label{detevaluated}
\end{align}
with the definitions
\begin{equation}
q_\pm^2=(E\pm\varepsilon)^2-(v_yk_y)^2-(v_zk_z)^2,\;\;w_\pm = \frac{W}{\hbar v_x}q_\pm.
\end{equation}

In the zero-offset limit $\varepsilon=0$ Eq.\ \eqref{detevaluated} reduces to the more compact expression
\begin{equation}
\left(\frac{v_z k_z}{q}\tan \frac{Wq}{\hbar v_x}\right)^2=1,\;\;q^2=E^2-(v_y k_y)^2-(v_z k_z)^2,\label{dispersionepszero}
\end{equation}
which is a squared Weiss equation \cite{Bov18,Bar17}.

\begin{figure}[tb]
\centerline{\includegraphics[width=1\linewidth]{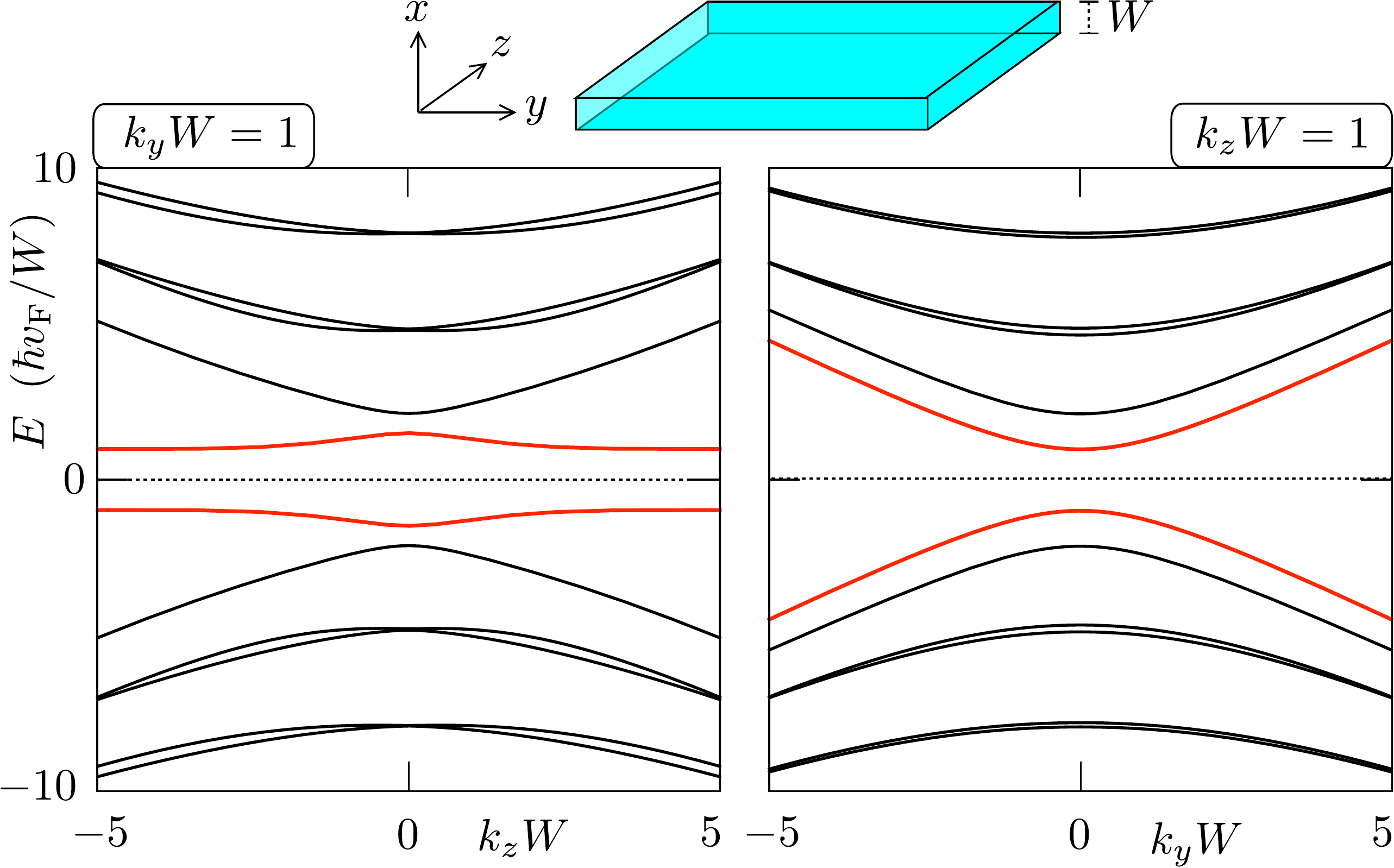}}
\caption{Dispersion relation $E(k_y,k_z)$ as a function of $k_z$ for fixed $k_y=1/W$ (left panel) and as a function of $k_y$ for fixed $k_z=1/W$ (right panel), calculated from Eq.\ \eqref{detevaluated} for $v_x=v_y=v_z\equiv v_{\rm F}$ and $\varepsilon=\hbar v_{\rm F}/W$. The surface states are indicated in red. The avoided crossings at $k_z=0$ become real crossings for $\varepsilon=0$.
}
\label{fig_dispersion}
\end{figure}

The dispersion relation $E(k_y,k_z)$ which follows from Eq.\ \eqref{detevaluated} is plotted in Fig.\ \ref{fig_dispersion}. The surface states (indicated in red) are nearly flat as function of $k_z$, so they propagate mainly in the $\pm y$ direction. In the limit $\varepsilon\rightarrow 0$ the bands cross at $k_z=0$, this crossing is removed by the energy offset.

\subsection{Fermi surface topology}
\label{sec_quasi2D}

\begin{figure}[tb]
\centerline{\includegraphics[width=1\linewidth]{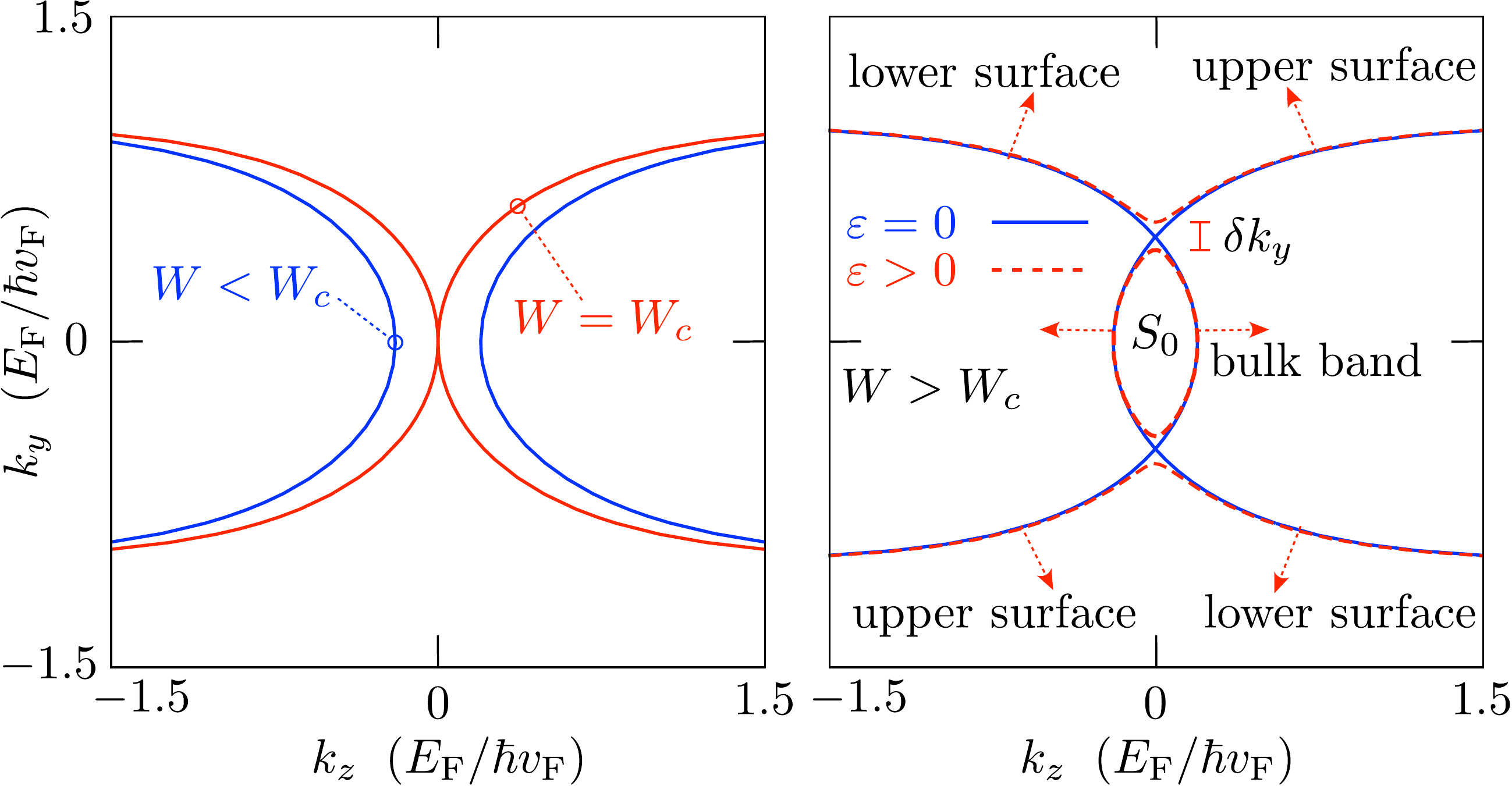}}
\caption{Solid curves: equi-energy contours $E(k_y,k_z)=E_{\rm F}$ for $\varepsilon=0$ at three values of $W$ (in units of $\hbar v_{\rm F}/E_{\rm F}$ with $E_{\rm F}>0$): $W=\pi/2=W_{\rm c}$ (red curve in left panel), $W=1.4<W_{\rm c}$ (blue curve in left panel), and $W=1.8>W_{\rm c}$ (blue curve in right panel). The calculations are based on Eq.\ \eqref{detevaluated} with $v_x=v_y=v_z\equiv v_{\rm F}$. The red dashed curve in the right panel shows the effect of a nonzero $\varepsilon=0.1\,E_{\rm F}$: The intersecting contours break up into two open and one closed contour, separated at $k_z=0$ by a gap $\delta k_y$.  The dotted arrows, perpendicular to the equi-energy contours, point into the direction of motion in real space. The assignment of the bands to the upper and lower surface is in accord with the time-reversal symmetry requirement that a band stays on the same surface when $(k_y,k_z)\mapsto -(k_y,k_z)$.
}
\label{fig_Fermisurface}
\end{figure}

The equi-energy contours $E(k_y,k_z)=E_{\rm F}$ are plotted in Fig.\ \ref{fig_Fermisurface} for several values of $W$. The topology of the Fermi surface changes at a critical width
\begin{equation}
W_{\rm c}=\frac{\pi}{2}\frac{\hbar v_{x}}{E_{\rm F}}+{\cal O}(\varepsilon).
\end{equation}
At $W=W_{\rm c}$ the surface bands from upper and lower surface touch at the Weyl point $k_y=k_z=0$, and for larger widths the upper and lower surface bands decouple from a bulk band, in the interior of the slab.

For $\varepsilon=0$ the surface and bulk bands intersect at $k_z=0$ when $W>W_{\rm c}$. The gap $\delta k_y$ which opens up for nonzero $\varepsilon$ is 
\begin{equation}
\delta k_y=\frac{4}{\pi v_y}|\varepsilon|+{\cal O}(\varepsilon^2),\;\;W>W_c.\label{deltakyresult}
\end{equation}
For later use we also record the area $S_0$ enclosed by the bulk band,
\begin{equation}
S_0=\tfrac{4}{3}\pi \sqrt 2 (W/W_c-1)^{3/2}k_{\rm F}^2+{\cal O}(W/W_c-1)^2+{\cal O}(\varepsilon),\label{S0result}
\end{equation}
where we have defined the 2D Fermi wave vector of the Weyl fermions via
\begin{equation}
E_{\rm F}=\hbar k_{\rm F}\sqrt{v_y v_z}.
\end{equation}

\section{Resonant tunneling between open and closed orbits in a magnetic field}
\label{sec_openclosed}

Upon application of a magnetic field $B$ in the $x$-direction, perpendicular to the slab, the Lorentz force causes a wave packet to drift along an equi-energy contour. Because $\dot{\bm{k}}=e\dot{\bm{r}}\times\bm{B}$ the orbit in real space is obtained from the orbit in momentum space by rotation over $\pi/2$ and rescaling by a factor $\hbar/eB=l_m^2$ (magnetic length squared).

\begin{figure}[tb]
\centerline{\includegraphics[width=1\linewidth]{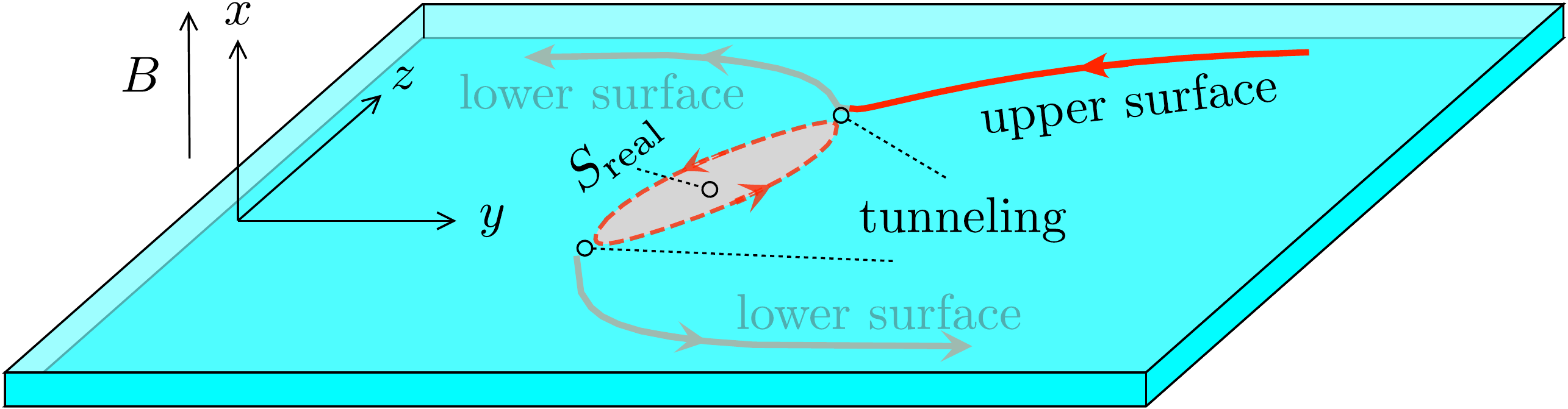}}
\caption{Electron orbits in a magnetic field perpendicular to the slab, following from the Fermi surface in Fig.\ \ref{fig_Fermisurface} ($W>W_c$, $\varepsilon>0$). The tunneling events (magnetic breakdown) between open and closed orbits are indicated. These happen with probability $T_{\rm MB}$ given by Eq.\ \eqref{TMBresult}. Backscattering of the open orbit via the closed orbit happens with probability ${\cal R}$ given by Eq.\ \eqref{Rresult}. The area $S_{\rm real}\propto 1/B^2$ of the closed orbit in real space determines the $1/B$ periodicity of the magnetoconductance oscillations via the resonance condition $BS_{\rm real}=nh/e$.
}
\label{fig_orbits}
\end{figure}

Inspection of Fig.\ \ref{fig_Fermisurface} shows that for $W>W_{c}$ closed orbits in the interior of the slab coexist with open orbits on the surface. The open and closed orbits are coupled via tunneling through a momentum gap $\delta k_y$ (magnetic breakdown \cite{Pip69,Kag83}), with tunnel probability $T_{\rm MB}=1-R_{\rm MB}$ given by the Landau-Zener formula
\begin{equation}
T_{\rm MB}=\exp(-B_c/B),\;\;B_c\simeq(\hbar/e)\delta k_y^2\simeq	(\hbar\varepsilon/ev_{\rm F})^2.\label{TMBresult}
\end{equation}
In the expression for the breakdown field $B_{\rm c}$ a numerical prefactor of order unity is omitted \cite{Kag83,Sta67}.

The real-space orbits are illustrated in Fig.\ \ref{fig_orbits}: An electron in a Fermi arc on the top surface switches to the bottom surface when the Fermi arc terminates at a Weyl point \cite{Pot14}. The direction of propagation (helicity) of the surface electron may change as a consequence of the magnetic breakdown, which couples a right-moving electron on the top surface to a left-moving electron on the bottom surface. This backscattering process occurs with reflection probability
\begin{equation}
{\cal R}=\left|\frac{T_{\rm MB}}{1-R_{\rm MB}e^{i\phi}}\right|^2=\frac{T_{\rm MB}^2}{T_{\rm MB}^2+4R_{\rm MB}\sin^2(\phi/2)}.\label{Rresult}
\end{equation}
The phase shift $\phi$ accumulated in one round trip along the closed orbit is determined by the enclosed area $S_0$ in momentum space,
\begin{equation}
\phi=S_0 l_m^2+2\pi\nu,\label{phiS0}
\end{equation}
with $\nu\in[0,1)$ a magnetic-field independent offset.

Resonant tunneling through the closed orbit, resulting in ${\cal R}=1$, happens when $\phi$ is an integer multiple of $2\pi$. We thus see that the resonances are periodic in $1/B$, with period
\begin{equation}
\Delta(1/B)=\frac{2\pi e}{\hbar S_0}\approx\frac{e}{h}(W/W_c-1)^{-3/2}k_{\rm F}^{-2}.\label{delta1B}
\end{equation}
(We have substituted the small-$\varepsilon$ expression \eqref{S0result} for $S_0$.)

The Shubnikov-de Haas (SdH) oscillations due to Landau level quantisation are also periodic in $1/B$. Their period is determined by the area $S_{\Sigma}\approx 2\pi k_{\rm F}/a_0$ in Fig.\ \ref{fig_chainnolabels}, hence
\begin{equation}
\Delta(1/B)_{\rm SdH}=\frac{2\pi e}{\hbar S_\Sigma}\approx\frac{ea_0}{\hbar k_{\rm F}}.\label{delta1BSdH}
\end{equation}
Comparison with Eq.\ \eqref{delta1B} shows that the period of the SdH oscillations is smaller than that of the magnetic breakdown oscillations by a factor $k_{\rm F}a_0(W/W_c-1)^{3/2}$, which is typically $\ll 1$. 

\section{Dispersive Landau bands}
\label{sec_dispersiveLL}

\begin{figure}[tb]
\centerline{\includegraphics[width=1\linewidth]{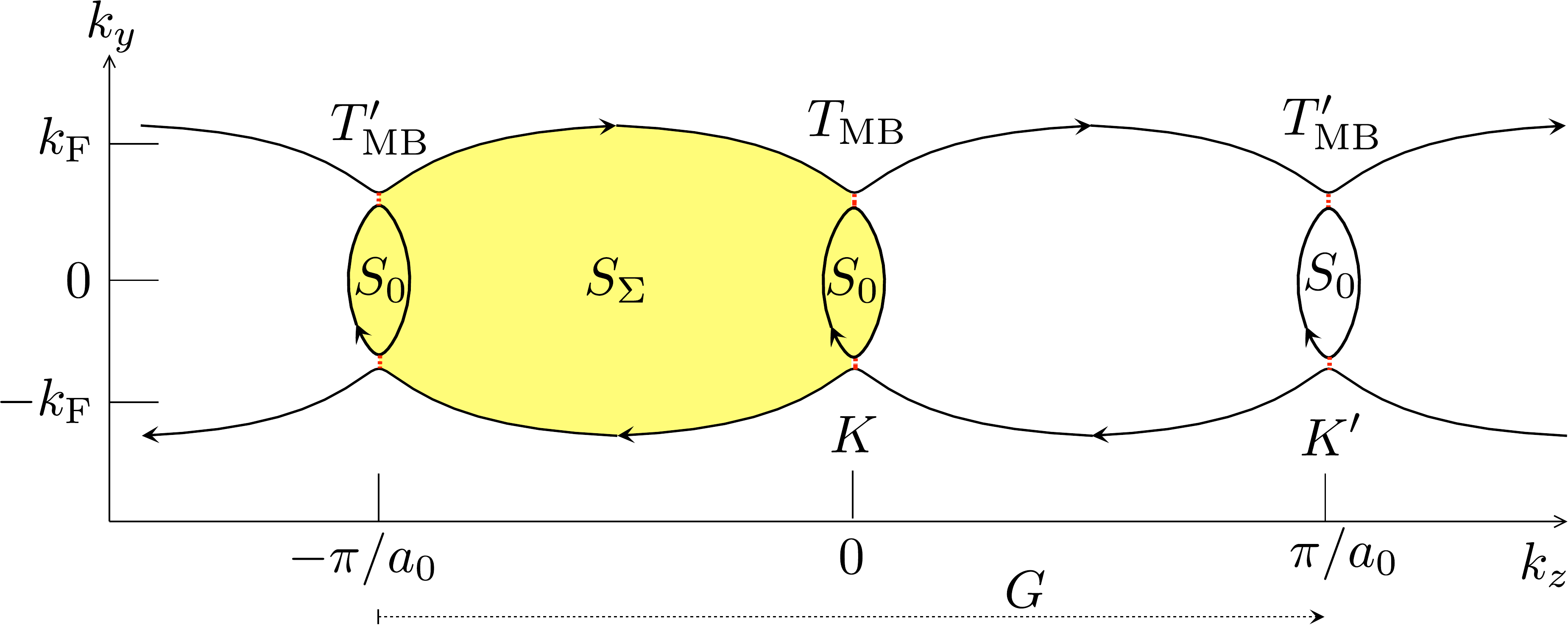}}
\caption{Equi-energy contours in the $k_y$--$k_z$ plane, showing open orbits coupled to closed orbits via magnetic breakdown (red dotted lines). The closed contours encircle Weyl points at $\bm{K}=(0,0)$ and $\bm{K}'=(0,\pi/a_0)$ --- periodically translated by the reciprocal lattice vector $\bm{G}=(0,2\pi/a_0)$. Arrows indicate the spectral flow in a perpendicular magnetic field. The large area $S_\Sigma$ (yellow) determines the spacing of the Landau bands, while the small area $S_0$ and the magnetic breakdown probabilities $T_{\rm MB},T'_{\rm MB}$ determine the band width.}
\label{fig_chainnolabels}
\end{figure}

Let us now discuss how magnetic breakdown converts the flat dispersionless Landau levels into dispersive bands. The mechanism crucially relies on the fact that the surface Fermi arcs in a Kramers-Weyl semimetal connect Weyl points at time-reversally invariant momenta. Consider two TRIM $\bm{K}$ and $\bm{K}'$ in the $(k_y,k_z)$ plane of the surface Brillouin zone. We choose $\bm{K}=(0,0)$ at the zone center and $\bm{K}'=(0,\pi/a_0)$ at the zone boundary, with $\bm{G}=(0,2\pi/a_0)$ a reciprocal lattice vector.

In the periodic zone scheme, the Weyl points can be repeated along the $k_z$-axis with period $2\pi/a_0$, to form an infinite one-dimensional chain (see Fig.\ \ref{fig_chainnolabels}). The perpendicular magnetic field $B$ induces a flow along this chain in momentum space, which in real space is oriented along the $y$-axis with period
\begin{equation}
{\cal L}=(2\pi/a_0)l_m^2=2\pi v_y/\omega_c,\;\;\omega_c=eBv_y a_0/\hbar.
\end{equation}
In the weak-field regime $l_m\gg a_0$ the period ${\cal L}$ of the magnetic-field induced superlattice is large compared to the period $a_0$ of the atomic lattice. We seek the band structure of the superlattice. 

We distinguish the Weyl points at $\bm{K}$ and $\bm{K}'$ by their different magnetic breakdown probabilities, denoted respectively by $T_{\rm MB}=1-R_{\rm MB}$ and $T'_{\rm MB}=1-R'_{\rm MB}$. We focus on the case that $T_{\rm MB}$ and $T'_{\rm MB}$ are close to unity and the areas $S_0$ and $S'_0$ of the closed orbits are the same --- this is the small-$\varepsilon$ regime in Eqs.\ \eqref{S0result} and \eqref{TMBresult}. (The more general case is treated in App.\ \ref{sec_spectrum}.)

The phase shift $\psi$ accumulated upon propagation from one Weyl point to the next is gauge dependent, we choose the Landau gauge $\bm{A}=(0,-Bz,0)$. For simplicity we ignore the curvature of the open orbits, approximating them by straight contours along the line $k_y=E/\hbar v_{y}$. The phase shift is then given by
\begin{equation}
\psi= \frac{E}{\hbar v_{y}}\frac{\pi}{a_0}l_m^2= \frac{\pi E}{\hbar\omega_c},
\end{equation}
the same for each segment of an open orbit connecting two Weyl points.

\begin{figure}[tb]
\centerline{\includegraphics[width=0.9\linewidth]{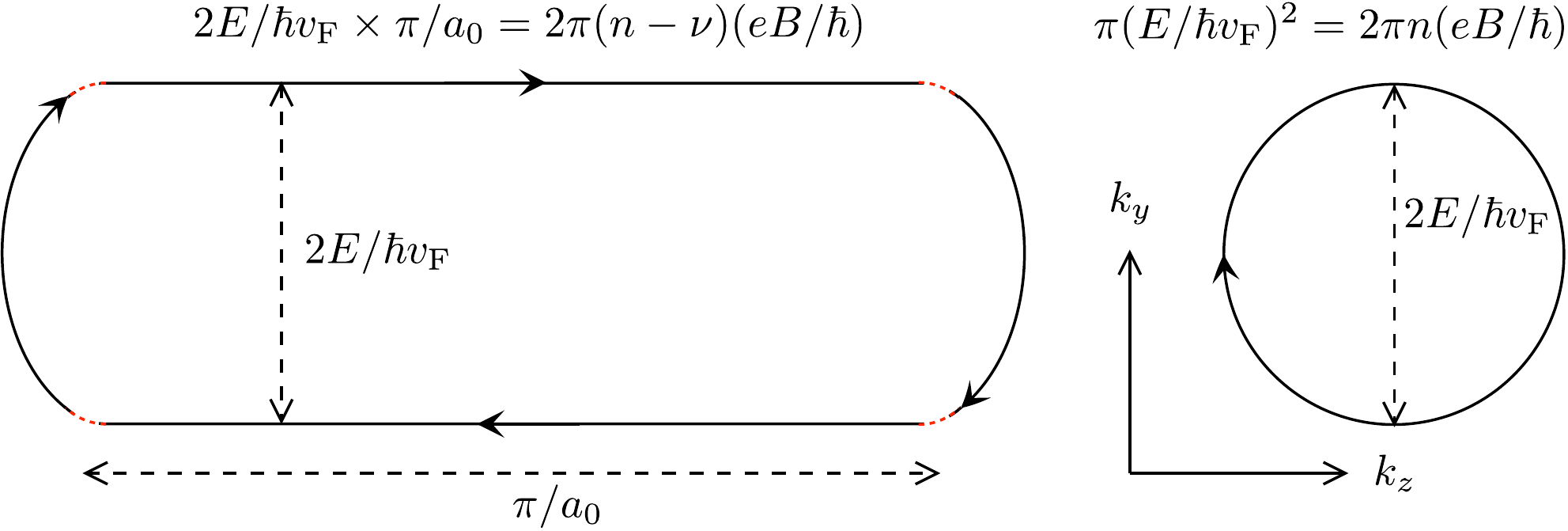}}
\caption{Equi-energy contours in the $k_y$--$k_z$ plane for surface Fermi arcs coupled by magnetic breakdown (left panel, schematic) and for the bulk cyclotron orbit of a Weyl fermion (right panel). The quantization condition for the enclosed area is indicated, to explain why the Landau level spacing is $\propto B$ for the Fermi arcs, while it is $\propto \sqrt{B}$ for the cyclotron orbit.
}
\label{fig_contours}
\end{figure}

The quantization condition for a Landau level at energy $E_n$ is $2\psi+\phi=2\pi n$, $n=1,2,\ldots$, which amounts to the quantization in units of $h/e$ of the magnetic flux through the real-space area $S_\Sigma l_m^4$. Since $S_\Sigma\gg S_0$ the Landau level spacing is governed by the energy dependence of $\psi$, 
\begin{equation}
E_{n+1}-E_n\approx \pi(d\psi/dE)^{-1}=\hbar\omega_c.
\end{equation}
The Landau level spacing increases $\propto B$ and not $\propto \sqrt B$, as one might have expected for massless electrons. The origin of the difference is explained in Fig.\ \ref{fig_contours}.

\begin{figure}[tb]
\centerline{\includegraphics[width=1\linewidth]{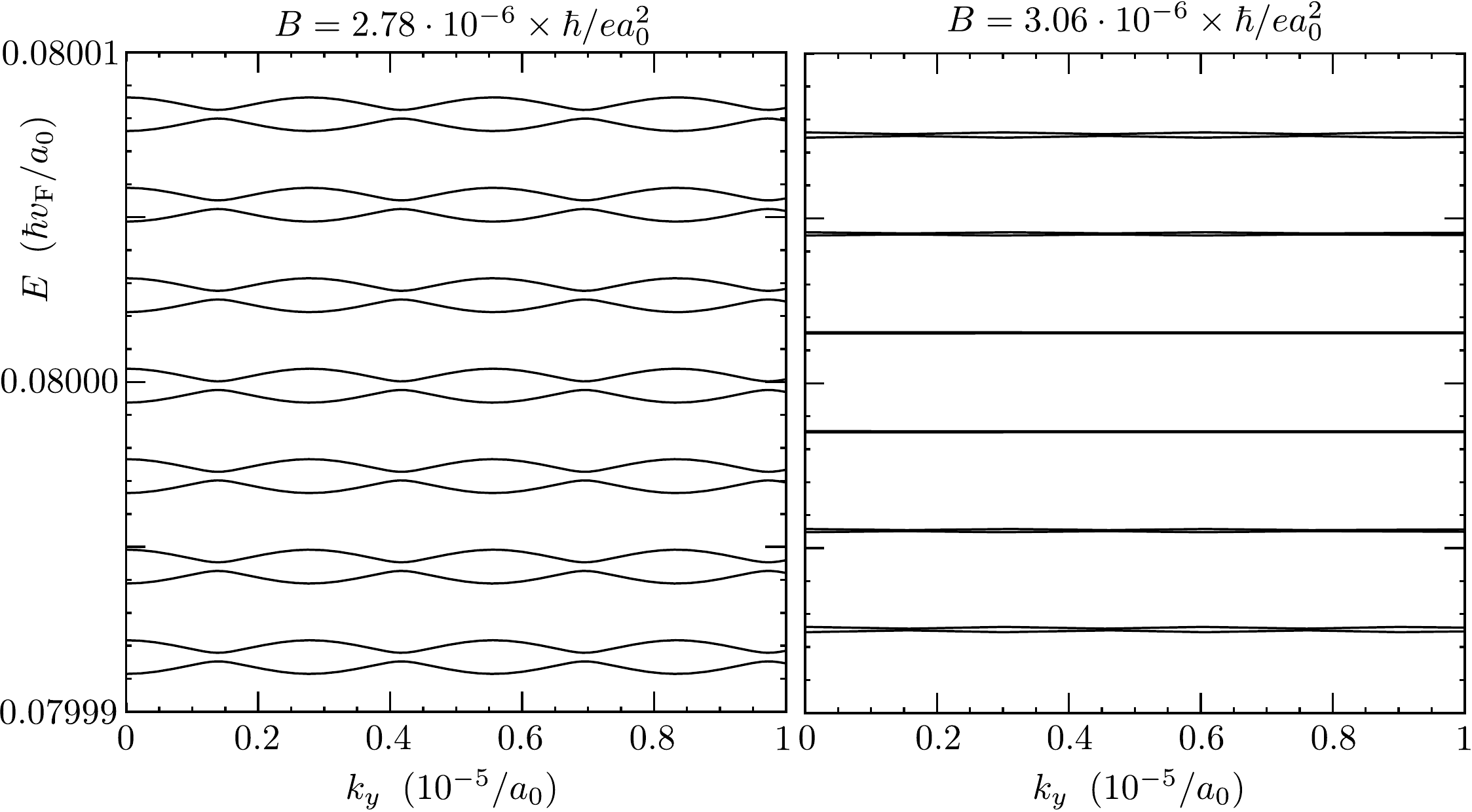}}
\caption{Dispersion relation of the slab in a perpendicular magnetic field $B$, calculated from Eqs.\ \eqref{Ttotalresulttr} and \eqref{trTtotalcosqL} (for $W=20\,a_0$, $T_{\rm MB}=0.85$, $T'_{\rm MB}=0.95$, $S_0$=$S'_0$, $\nu=0$). In the left panel $B$ is chosen such that the phase $\phi$ accumulated by a closed orbit at $E=0.08\,\hbar v_{\rm F}/a_0$ equals $11\pi$, in the right panel $\phi=10\pi$. When $\phi$ is an integer multiple of $2\pi$ the magnetic breakdown is resonant, all orbits are closed and the Landau bands are dispersionless. When $\phi$ is a half-integer multiple of $2\pi$ the magnetic breakdown is suppressed and the Landau bands acquire a dispersion from the open orbits.
}
\label{fig_oscillation}
\end{figure}

The Landau levels are flat when $T_{\rm MB}=T'_{\rm MB}=1$, so that there are no open orbits. The open orbits introduce a dispersion along $k_y$, see Fig.\ \ref{fig_oscillation}. Full expressions are given in App.\ \ref{sec_spectrum}. For $R_{\rm MB},R'_{\rm MB}\ll 1$ and $S_0=S'_0$ we have the dispersion
\begin{align}
&E(k_y)=(n-\nu)\hbar\omega_c\pm (\hbar\omega_c/\pi)\sin(\phi/2)\nonumber\\
&\qquad\times\bigl(R_{\rm MB}+R'_{\rm MB}+2\sqrt{R_{\rm MB}R'_{\rm MB}}\cos k_y{\cal L}\bigr)^{1/2},\label{dispersive_LL}
\end{align}
where the phase $\phi$ is to be evaluated at $E=(n-\nu)\hbar\omega_c$.

Each Landau level is split into two subbands having the same band width
\begin{align}
&| E(0)-E(\pi/{\cal L})|=\nonumber\\
&\qquad 2(\hbar \omega_c/\pi)|\sin(\phi/2)|\min(\sqrt{R}_{\rm MB},\sqrt{R'}_{\rm MB}).
\end{align}
The band width oscillates periodically in $1/B$ with period \eqref{delta1B}.

\section{Magnetoconductance oscillations}
\label{sec_magneto}

The dispersive Landau bands leave observable signatures in electrical conduction, in the form of magnetoconductance oscillations due to the resonant coupling of closed and open orbits. These have been previously studied when the open orbits are caused by an electrostatic superlattice \cite{Ger89,Win89,Bee89,Str90,Gvo07}. We apply that theory to our setting.

From the dispersion relation \eqref{dispersive_LL} we calculate the square of the group velocity ${\cal V}=\partial E/\hbar\partial k_y$, averaged over the Landau band,
\begin{align}
\langle {\cal V}^2\rangle&=\frac{\cal L}{2\pi}\int_0^{2\pi/{\cal L}}\left(\frac{dE(k_y)}{\hbar dk_y}\,\right)^2 dk_y \nonumber\\
&=2v_{y}^2\sin^2(\phi/2)\min(R_{\rm MB},R'_{\rm MB}).
\end{align}

For weak impurity scattering, scattering rate $1/\tau_{\rm imp}\ll\omega_c$, the effective diffusion coefficient \cite{Gvo07},
\begin{equation}
D_{\rm eff}=\tau_{\rm imp}\langle {\cal V}^2\rangle,
\end{equation}
and the 2D density of states $N_{\rm 2D}=(\pi\hbar v_y a_0)^{-1}$ of the Landau band, determine the oscillatory contribution $\delta\sigma_{yy}$ to the longitudinal conductivity via the Drude formula for a 2D electron gas,
\begin{align}
\delta\sigma_{yy}&=e^2N_{\rm 2D} D_{\rm eff}\nonumber\\
&=\frac{4e^2}{h}\frac{v_y\tau_{\rm imp}}{a_0}\,\sin^2(\phi/2)\min(R_{\rm MB},R'_{\rm MB}).
\end{align}

The magnetoconductance oscillations due to magnetic breakdown (MB) coexist with the Shubnikov-de Haas (SdH) oscillations due to Landau level quantization. Both are periodic in $1/B$, but with very different period, see Eqs.\ \eqref{delta1B} and \eqref{delta1BSdH}. 

The difference in period causes a different temperature dependence of the magnetoconductance oscillations. A conductance measurement at temperature $T$ corresponds to an energy average over a range $\Delta E\approx 4k_{\rm B}T$ (being the full-width-at-half-maximum of the derivative of the Fermi-Dirac distribution). The oscillations become unobservable when the energy average changes the area $S_0$ or $S_\Sigma$ by more than $\pi/l_m^2$. This results in different characteristic energy or temperature scales,
\begin{subequations}
\label{DeltaESdHMB}
\begin{align}
&\Delta E_{\rm SdH}=\frac{\pi}{l_m^2}\left(\frac{\partial S_\Sigma}{\partial E}\right)^{-1}\simeq\tfrac{1}{2}\hbar\omega_c,\\
&\Delta E_{\rm MB}=\frac{\pi}{l_m^2}\left(\frac{\partial S_0}{\partial E}\right)^{-1}\simeq\tfrac{1}{4}\sqrt 2(W/W_c-1)^{-1/2}\frac{\hbar\omega_c}{k_{\rm F}a_0}.
\end{align}
\end{subequations}
(In the second equation we took $W/W_c\gtrsim 1$.) For $k_{\rm F}a_0\ll 1$ and $W/W_c$ close to unity we may have $\Delta E_{\rm SdH}\ll \Delta E_{\rm MB}$, so there is an intermediate temperature regime $\Delta E_{\rm SdH}\lesssim 4k_{\rm B}T\lesssim \Delta E_{\rm MB}$ where the Shubnikov-de Haas oscillations are suppressed while the magnetic breakdown oscillations remain.

\section{Tight-binding model on a cubic lattice}
\label{sec-tbmodel}

We have tested the analytical calculations from the previous sections numerically, on a tight-binding model of a Kramers-Weyl semimetal \cite{Cha18}. In this section we describe the model, results are presented in the next section.

\subsection{Hamiltonian}
\label{sec_Hamiltonian}

We take a simple cubic lattice (lattice constant $a$, one atom per unit cell), when the nearest-neighbor hopping terms are the same in each direction $\alpha\in\{x,y,z\}$. There are two terms to consider, a spin-independent term $\propto t_0$ that is even in momentum and a spin-orbit coupling term $\propto t_1 \sigma_\alpha$ that is odd in momentum,
\begin{equation}
H=t_0\sum_{\alpha}\cos (k_\alpha a)+t_1\sum_\alpha \sigma_\alpha\sin (k_\alpha a)-t_0.
\label{tbham}
\end{equation}
The offset is arbitrarily fixed at $-t_0$.

There are 8 Weyl points (momenta $\bm{k}$ in the Brillouin zone of a linear dispersion), located at $k_x,k_y,k_z\in\{0,\pi\}$ modulo $2\pi$. The Weyl points at $(k_x,k_y,k_z)=(0,0,0),(\pi,\pi,0),(\pi,0,\pi),(0,\pi,\pi)$ have positive chirality and those at $(\pi,\pi,\pi),(\pi,0,0),(0,\pi,0),(0,0,\pi)$ have negative chirality \cite{Cha18}.

The geometry is a slab, with a normal $\hat{n}$ in the $x$--$z$ plane at an angle $\phi$ with the $x$-axis (so the normal is rotated by $\phi$ around the $y$-axis). The boundaries of the slab are constructed by removing all sites at $x<0$ and $x>W$. In the rotated basis aligned with the normal to the slab one has
\begin{equation}
\begin{pmatrix}
k'_x\\
k'_z
\end{pmatrix}=\begin{pmatrix}
\cos\phi&\sin\phi\\-\sin\phi&\cos\phi
\end{pmatrix}\begin{pmatrix}
k_x\\
k_z
\end{pmatrix},\;\;k'_y=k_y.
\end{equation}
We will work in this rotated basis and for ease of notation omit the prime, writing $k_x$ or $k_{\perp}$ for the momentum component perpendicular to the slab and $(k_y,k_z)=\bm{k}_\parallel$ for the parallel momenta.

\subsection{Folded Brillouin zone}

\begin{figure}[tb]
\centerline{\includegraphics[width=0.9\linewidth]{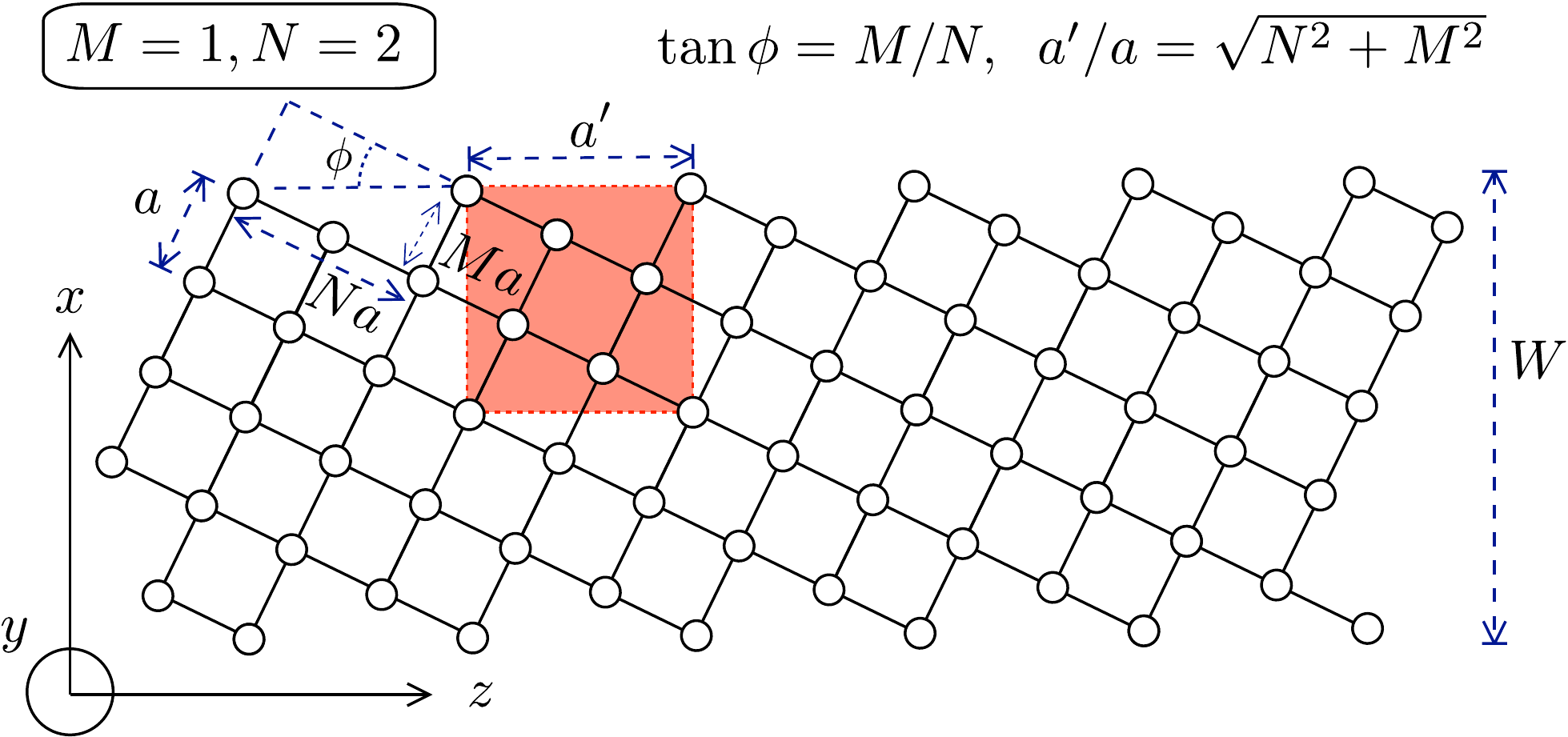}}
\caption{Slice at $y=0$ through the cubic lattice, rotated around the $y$-axis by an angle $\phi=\arctan(M/N)$ with $M=1$, $N=2$. The enlarged unit cell (red square), parallel to a lattice termination at $x=0$ and $x=W$, has volume $a'\times a'\times a=(N^2+M^2)a^3$.
}
\label{fig_rotation}
\end{figure}

\begin{figure}[tb]
\centerline{\includegraphics[width=1\linewidth]{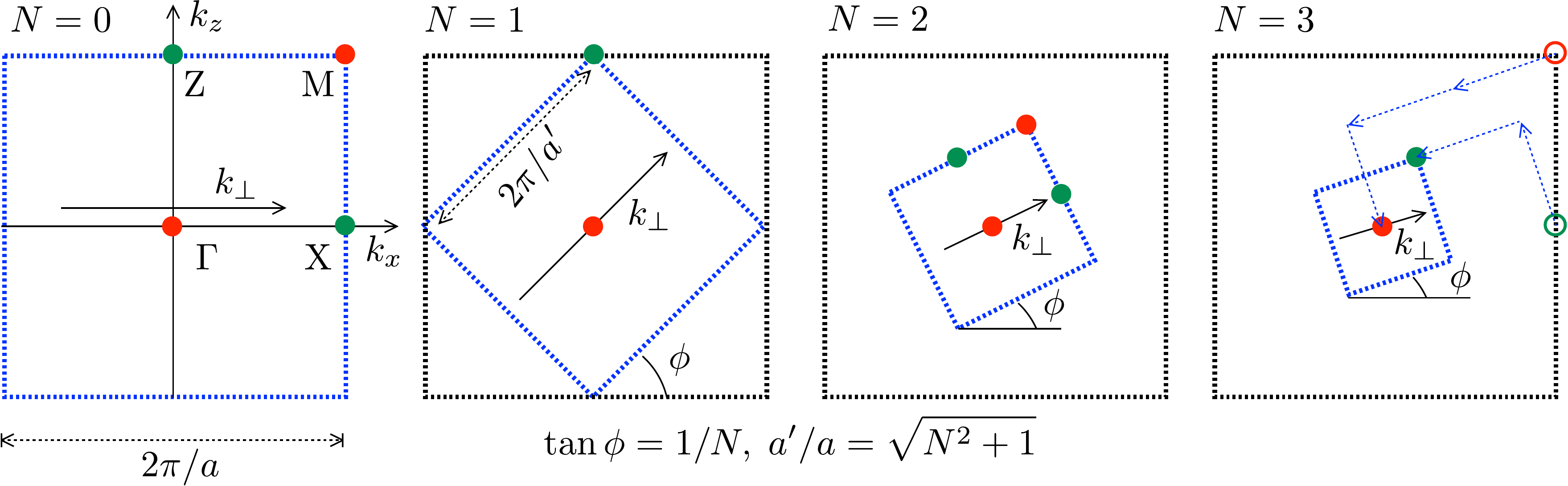}}
\caption{Slice at $k_y=0$ through the Brillouin zone of the rotated cubic lattice, for rotation angles $\phi=\arctan(M/N)$ with $M=1$, $N=0,1,2,3$. Weyl points of opposite chirality are marked by a green or red dot. The panel for $N=3$ shows how translation by reciprocal lattice vectors (blue arrows) folds two Weyl points onto each other.
}
\label{fig_Brillouin}
\end{figure}

The termination of the lattice in the slab geometry breaks the translation invariance in the perpendicular $x$-direction as well as in the $z$-direction parallel to the surface. If the rotation angle $\phi\in(0,\pi/2]$ is chosen such that $\tan\phi=M/N$ is a rational number ($M$ and $N$ being coprime integers), the translational invariance in the $z$-direction is restored with a larger lattice constant $a'=a\sqrt{N^2+M^2}$, see Fig.\ \ref{fig_rotation}. There are then $N^2+M^2$ atoms in a unit cell.

In reciprocal space the enlarged unit cell folds the Brillouin zone. Relative to the original Brillouin zone the folded Brillouin zone is rotated by an angle $\phi$ around the $y$-axis and scaled by a factor $(N^2+M^2)^{-1/2}$ in the $x$ and $z$-directions, see Fig.\ \ref{fig_Brillouin}. The reciprocal lattice vectors in the rotated basis are
\begin{equation}
\bm{e}_x=(2\pi/a')\hat{x},\;\;\bm{e}_y=(2\pi/a)\hat{y},\;\;
\bm{e}_z=(2\pi/a')\hat{z}.
\end{equation}

The corner in the $k_y=0$ plane of the original Brillouin zone (the M point) has coordinates 
\[
\frac{\pi}{a}(\cos\phi+\sin\phi,\cos\phi-\sin\phi,0)=\frac{\pi}{a'}(N+M,N-M,0)
\]
in the rotated lattice. Upon translation over a reciprocal lattice vector this is folded onto the center of the Brillouin zone (the $\Gamma$ point) when $N+M$ is an even integer, while it remains at a corner for $N+M$ odd. The midpoints of a zone boundary, the X and Z points, are folded similarly, as summarized by
\[
\begin{split}
&\text{M}\mapsto \Gamma, \;\;\Gamma\mapsto\Gamma,\;\; \text{X}\mapsto \text{M}, \;\; \text{Z}\mapsto \text{M},\;\;\text{for}\;\; N+M\;\;\text{even},\\
&\text{M}\mapsto \text{M}, \;\;\Gamma\mapsto\Gamma,\;\; \text{X}\mapsto \text{X}, \;\; \text{Z}\mapsto \text{Z},\;\;\text{for}\;\; N+M\;\;\text{odd}.
\end{split}
\]

Since the Weyl points at $\Gamma$ and M have the same chirality, for $N+M$ even we are in the situation that the surface of the slab couples Weyl points of the same chirality --- which is required for surface Fermi arcs to appear (see Sec.\ \ref{sec_Hbc}). For $N+M$ odd, in contrast, the Weyl points at the $\Gamma$ and X points of opposite chirality are coupled by the surface, since these line up along the $k_\perp$ axis. Then surface Fermi arcs will not appear. In App.\ \ref{app_chirality} we present a general analysis, for arbitrary Bravais lattices, that determines which lattice terminations support Fermi arcs and which do not.

\section{Tight-binding model results}
\label{sec_numerics}

We present results for $M=N=1$, corresponding to a $\phi=\pi/4$ rotation of the lattice around the $y$-axis. The folded and rotated Brillouin zone has a pair of Weyl points of $+$ chirality at $\bm{K}=(0,0,0)$ and a second pair of $-$ chirality at $\bm{K}'=(\pi/a',0,\pi/a')$ in the rotated coordinates (see Fig.\ \ref{fig_Brillouin}, second panel, with $a'=a\sqrt 2$). There is a second pair translated by $k_y=\pi/a$.

Each Weyl point supports a pair of Weyl cones of the same chirality, folded onto each other in the first Brillouin zone. The Weyl cones at $\bm{K}$ have energy offset $\varepsilon=|2t_0|$, while those at $\bm{K}'$ have  $\varepsilon'=0$. We may adjust the offset by adding a rotational symmetry breaking term $\delta H=\delta t_0\cos k_z a$ to the tight-binding Hamiltonian \eqref{tbham}. This changes the offsets into
\begin{equation}
\varepsilon=|2t_0+\delta t_0|,\;\;\varepsilon'=|\delta t_0|.
\end{equation}

\begin{figure}[tb]
\includegraphics[width=1\linewidth]{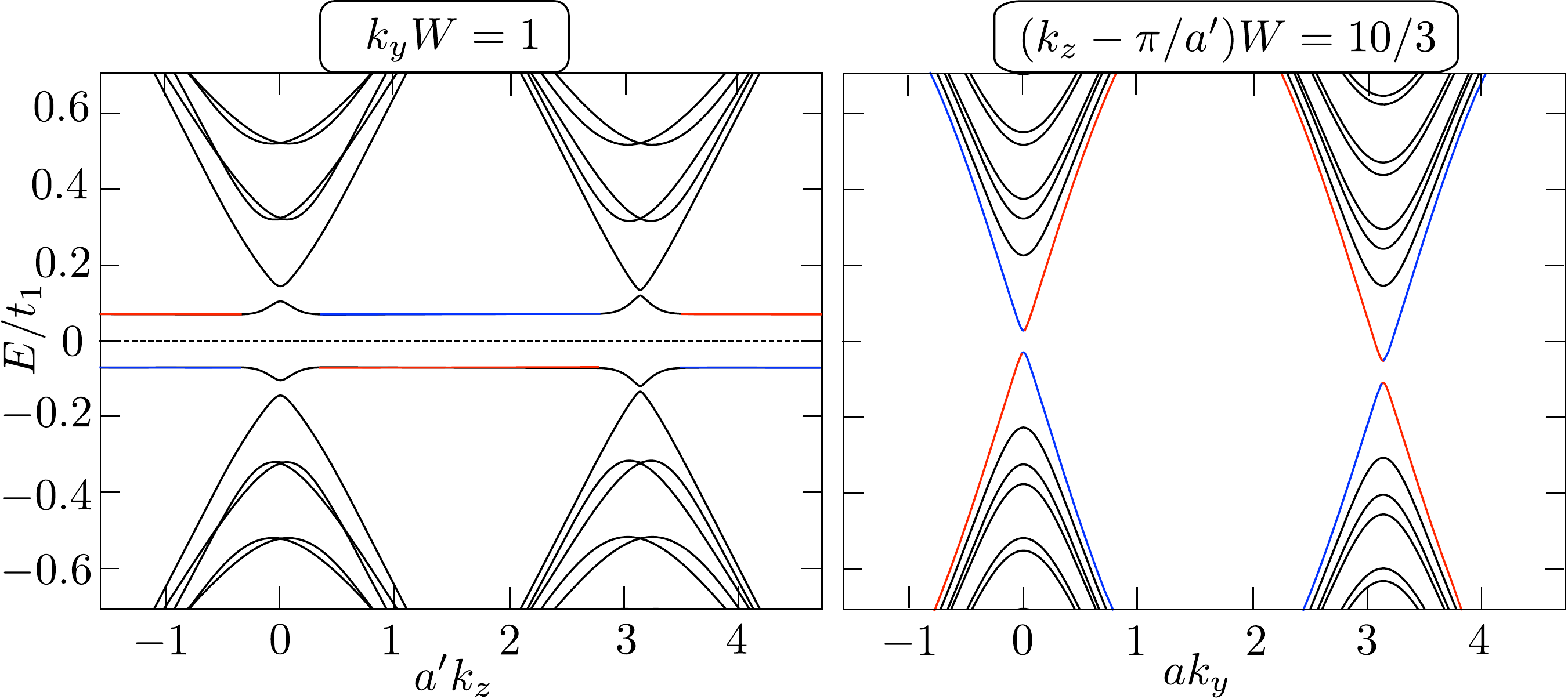}
\caption{Dispersion relations of a slab (thickness $W=10\sqrt{2}\,a$ in the $x$-direction, infinitely extended in the $y$--$z$ plane) in zero magnetic field. The plots are calculated from the tight-binding model of Sec.\ \ref{sec_numerics} (with $t_0=0.04\,t_1$, $\delta t_0=-0.02\,t_1$, corresponding to $\varepsilon=0.06\,t_1$, $\varepsilon'=0.02\,t_1$). The left and right panels show the dispersion as a function of $k_z$ and $k_y$, respectively. The curves are colored according to the electron density on the surfaces: red for the bottom surface, blue for the top surface, with bulk states appearing black. 
}
\label{fig_tb_dispersion}
\end{figure}

\begin{figure}[tb]
\includegraphics[width=1\linewidth]{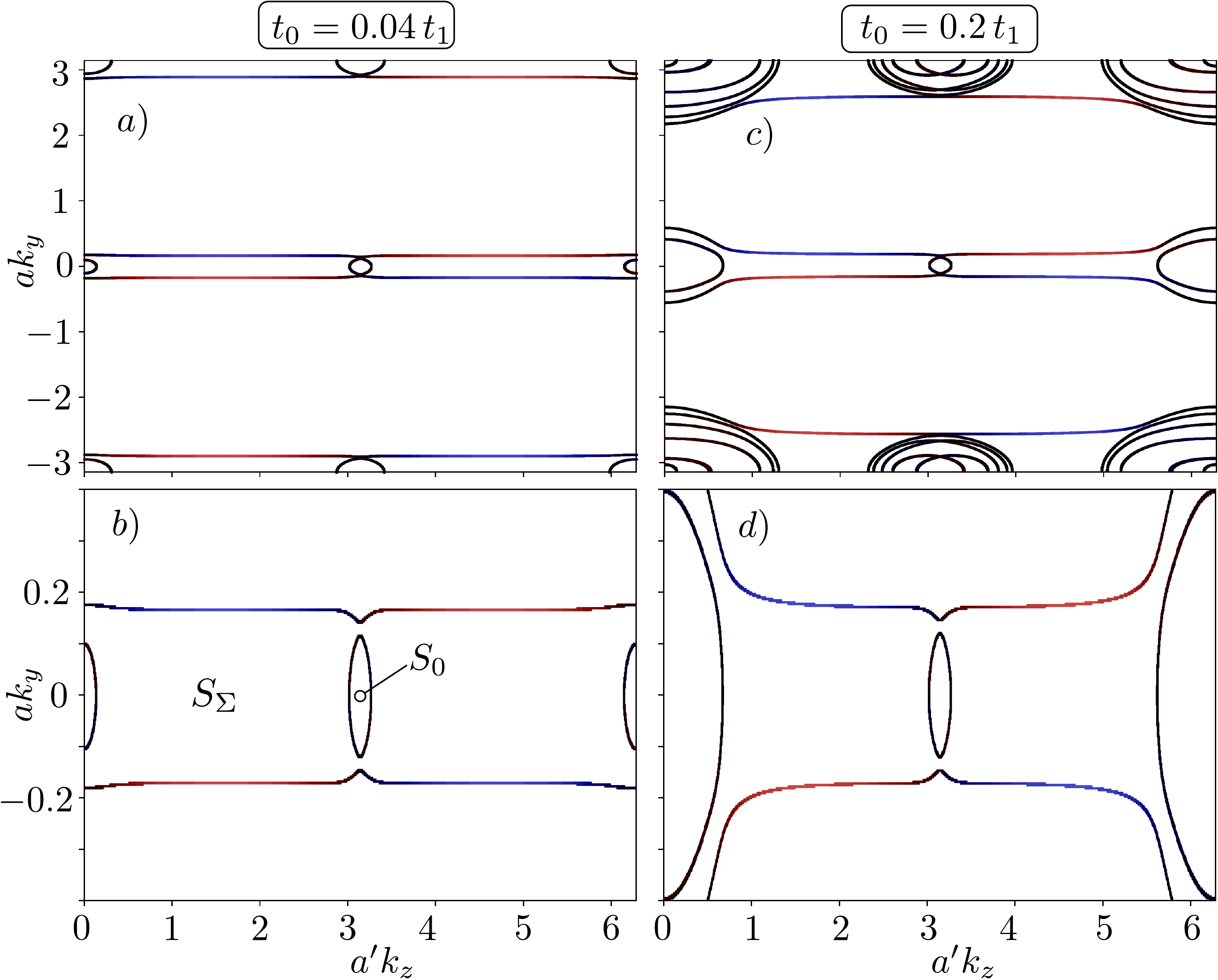}
\caption{Panels \textit{a} (full Brillouin zone) and \textit{b} (zoom-in near $k_y=0$) show equi-energy contours at $E=0.167\,t_1$ (when $W\approx 1.5\,W_c$), for the same system as in Fig.\ \ref{fig_tb_dispersion}. In panels \textit{c} and \textit{d} the spin-independent hopping term $t_0$ is increased by a factor 5 (at the same $\delta t_0=-0.02\,t_1$).
}
\label{fig_tb_fermisurface}
\end{figure}

In Fig.\ref{fig_tb_dispersion} we show how the Fermi arcs appear in the dispersion relation connecting the Weyl cones at $k_z=0$ and $k_z=\pi/a'$. This figure extends the local description near a Weyl cone from Fig.\ \ref{fig_dispersion} to the entire Brillouin zone. The corresponding equi-energy contours are presented in Fig.\ \ref{fig_tb_fermisurface}. Increasing the spin-independent hopping term $t_0$ introduces more bands, but the qualitative picture near the center of the Brillouin zone remains the same as in Fig.\ \ref{fig_Fermisurface} for $W>W_c$.

\begin{figure*}[tb]
\includegraphics[width=1\linewidth]{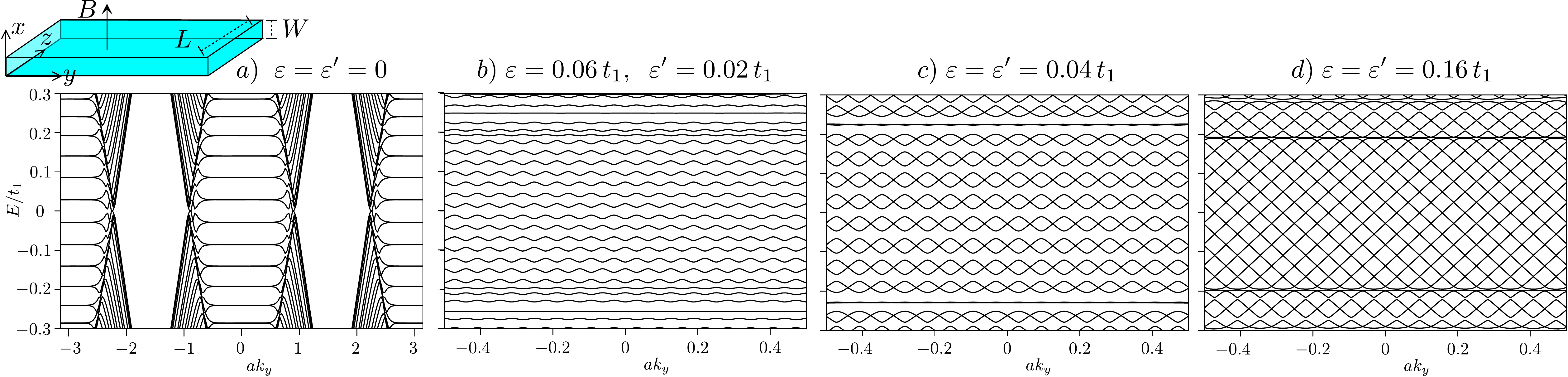}
\caption{Dispersion relation of a strip (cross-section $W\times L$ with $W=10a'$ and $L=30a'$) in a perpendicular magnetic field $B=0.00707\,(h/ea^2)$ (magnetic length $l_m=4.74\,a$). The four panels correspond to $t_0/t_1,\delta t_0/t_1$ equal to $0,0$ (panel \textit{a}), $0.04,-0.02$ (panel \textit{b}), $0.04,-0.04$ (panel \textit{c}), $0.16,-0.16$ (panel \textit{d}). The surface Fermi arcs near $k_y=0$ form closed orbits in panel \textit{a}, producing flat Landau levels, while in panel \textit{d} they form open orbits with the same linear dispersion as in zero field. Panels \textit{b,c} show an intermediate regime where magnetic breakdown between closed and open orbits produces Landau bands with an oscillatory dispersion.
}
\label{fig_tb_dispLL}
\end{figure*}

The effect on the dispersion of a magnetic field $B$, perpendicular to the slab, is shown in Fig.\ \ref{fig_tb_dispLL} (see also App.\ \ref{sec_landauoffset}). The field was incorporated in the tight-binding model via the Peierls substitution in the gauge $\bm{A}=(0,-Bz,0)$, with coordinate $z$ restricted to $|z|<L/2$. Translational invariance in the $y$-direction is maintained, so we have a one-dimensional dispersion $E(k_y)$. The boundaries of the system at $z=\pm L/2$ introduce edge modes, which are visible in panel \textit{a} as linearly dispersing modes near $k_y=\pm \tfrac{1}{2}L/l_m^{2}$ (modulo $\pi/a$). Panels \textit{b,c,d} focus on the region near $k_y=0$, where these edge effects can be neglected. The effect on the dispersion of a variation in $\varepsilon$ and $\varepsilon'$ is qualitatively similar to that obtained from the analytical solution of the continuum model, compare the four panels of Fig.\ \ref{fig_tb_dispLL} with the corresponding panels in Fig.\ \ref{fig_fourplots}.

\begin{figure}[tb]
\includegraphics[width=0.8\linewidth]{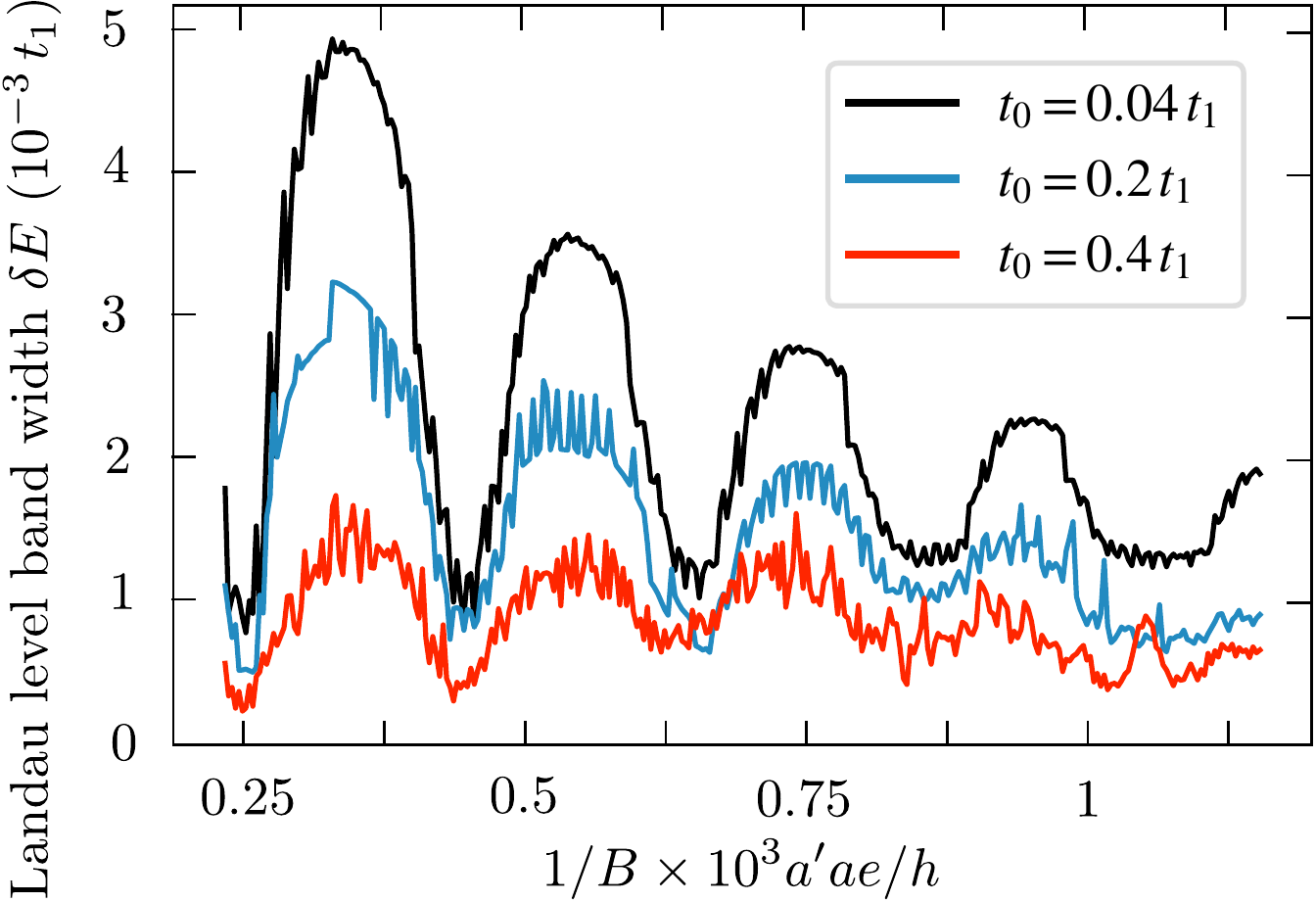}
\caption{Band width of the Landau levels versus inverse of magnetic field for $W=10a'$, $L=500a'$, $\delta t_0=-0.02\,t_1$ and three different values of $t_0$. The band widths are averaged over an energy window $\Delta E=0.004\,t_1$ around the Fermi energy $E_{\rm F}=0.167\,t_1$. The rapid Shubnikov-de Haas oscillations are averaged out, only the slow oscillations due to magnetic breakdown persist.}
\label{fig_tb_bandwidth}
\end{figure}

\begin{figure}[tb]
\includegraphics[width=0.8\linewidth]{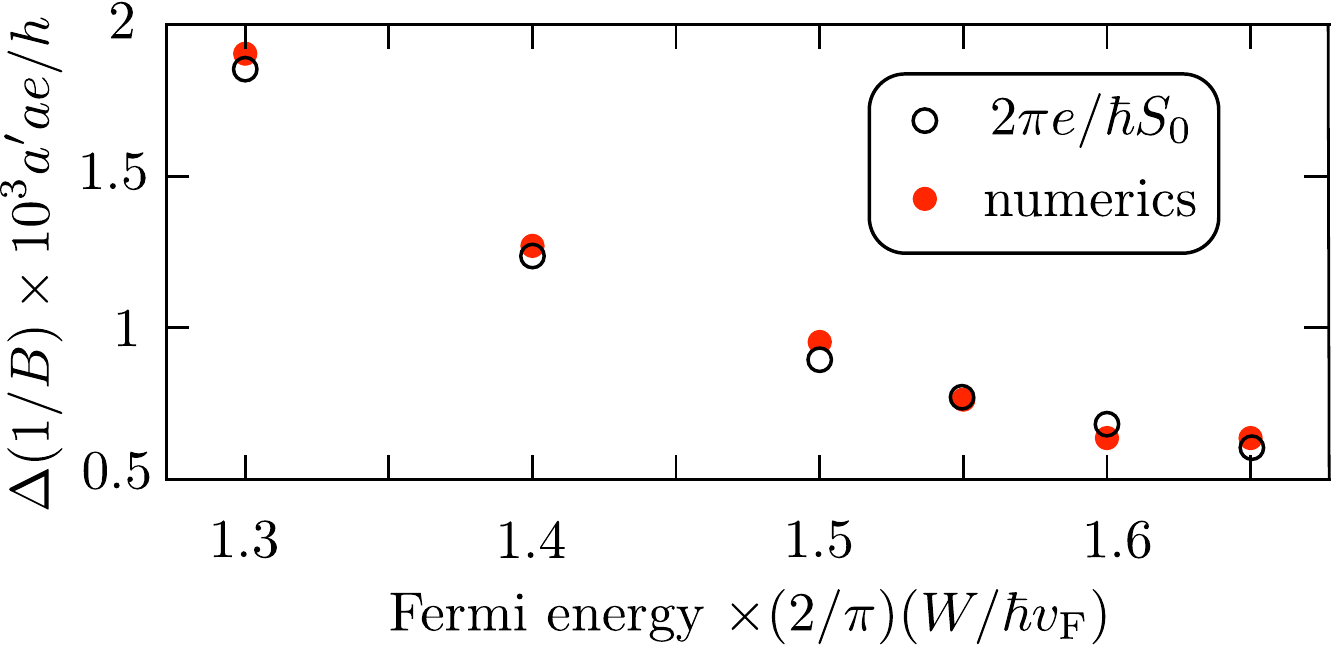}
\caption{Periodicity in $1/B$ of the Landau band width oscillations as a function of the Fermi energy, for $W=10a'$, $L=500a'$, $t_0=0.04\,t_1$, and $\delta t_0=-0.02\,t_1$. The filled data points are obtained numerically from the Landau band spectrum, similarly to the data shown for one particular $E_{\rm F}$ in Fig.\ \ref{fig_tb_bandwidth}. The open circles are calculated from the area $S_0$ of the closed orbit in momentum space (as indicated in Fig.\ \ref{fig_tb_fermisurface}\textit{b}), using the formula $\Delta(1/B)=2\pi e/\hbar S_0$.}
\label{fig_tb_oscperiod}
\end{figure}

The width $\delta E$ of the dispersive Landau bands (from maximum to minimum energy) is plotted as a function of $1/B$ in Fig.\ \ref{fig_tb_bandwidth} and the periodicity $\Delta(1/B)$ is compared with the predicted Eq.\ \eqref{delta1B} in Fig.\ \ref{fig_tb_oscperiod}. To remove the rapid Shubnikov-De Haas (SdH) oscillations we averaged over an energy interval $\Delta E$ around $E_{\rm F}$. This corresponds to a thermal average at effective temperature $T_{\rm eff}=\Delta E/4k_{\rm B}$. From Eq.\ \eqref{DeltaESdHMB}, with $k_{\rm F}a\approx 0.2$, $W/W_c\approx 1.5$, we estimate that the characteristic energy scale at which the oscillations average out is five times smaller for the SdH oscillations than for the oscillations due to magnetic breakdown, consistent with what we see in the numerics.

\section{Conclusion}
\label{sec_conclude}

In conclusion, we have shown that Kramers-Weyl fermions (massless fermions near time-reversally invariant momenta) confined to a thin slab have a fundamentally different Landau level spectrum than generic massless electrons: The Landau levels are not flat but broadened with a band width that oscillates periodically in $1/B$. The origin of the dispersion is magnetic breakdown at Weyl points, which couples open orbits from surface Fermi arcs to closed orbits in the interior of the slab.

The band width oscillations are observable as a slow modulation of the conductance with magnetic field, on which the rapid Shubnikov-de Haas oscillations are superimposed. The periodicities are widely separated because the quantized areas in the Brillouin zone are very different (compare the areas $S_0$ and $S_\Sigma$ in Fig.\ \ref{fig_chainnolabels}). This is a robust feature of the band structure of a Kramers-Weyl semimetal, as illustrated in the model calculation of Fig.\ \ref{fig_tb_fermisurface}. Since generic Weyl fermions have only the Shubnikov-de Haas oscillations, the observation of two distinct periodicities in the magnetoconductance would provide for a unique signature of Kramers-Weyl fermions.

The dispersive Landau band is interpreted as the band structure of a one-dimensional superlattice of magnetic breakdown centra, separated in real space by a distance ${\cal L}=(eBa_0/h)^{-1}$ --- which in weak fields is much larger than the atomic lattice constant $a_0$. Such a magnetic breakdown lattice has been studied in the past for massive electrons \cite{Kag83}, the Kramers-Weyl semimetals would provide an opportunity to investigate their properties for massless electrons.

\acknowledgments

The tight-binding model calculations were performed using the Kwant code \cite{kwant}. This project has received funding from the Netherlands Organization for Scientific Research (NWO/OCW) and from the European Research Council (ERC) under the European Union's Horizon 2020 research and innovation programme.

\appendix

\section{Coupling of time-reversally invariant momenta by the boundary}
\label{sec_proof}

The derivation of the boundary condition for Kramers-Weyl fermions in Sec.\ \ref{sec_Hbc} relies on pairwise coupling of Weyl cones at a TRIM by the boundary. Let us demonstrate that this is indeed what happens.

Consider a 3D Bravais lattice and its Brillouin zone. A time-reversally-invariant momentum (TRIM) is by definition a momentum $\bm{K}$ such that
$\bm{K} = -\bm{K}+\bm{G}$ with $\bm{G}$ a reciprocal lattice vector, or equivalently, $\bm{K}=\frac{1}{2}\bm{G}$. Now consider the restriction of the lattice to $x>0$, by removing all lattice points at $x<0$. Assume that the restricted lattice is still periodic in the $y$--$z$ plane, with an enlarged unit cell. Fig.\ \ref{fig_rotation} shows an example for a cubic lattice.

The enlarged unit cell will correspond to a reduced Brillouin zone, with a new set of reciprocal lattice vectors $\bm{\tilde G}$. The original set $\bm{K}_1,\bm{K}_2,\bm{K}_3,\ldots$ of TRIM is folded onto a new set $\bm{\tilde K}_1,\bm{\tilde K}_2,\bm{\tilde K}_3,\ldots $ in the reduced Brillouin zone. The folding may introduce degeneracies, such that two different $\bm{K}$'s are folded onto the same $\bm{\tilde K}$. The statement to prove is this:
\begin{itemize}
\item Each TRIM $\bm{\tilde{K}}$ in the folded Brillouin zone is either degenerate (because two $\bm{K}$'s were folded onto the same $\bm{\tilde{K}}$), or there is a second TRIM $\bm{\tilde K}'$ along the $k_x$-axis. 
\end{itemize}
Fig.\ \ref{fig_Brillouin} illustrates that this statement is true for the cubic lattice. We wish to prove that it holds for any Bravais lattice.

Enlargement of the unit cell changes the primitive lattice vectors from $\bm{a}_1,\bm{a}_2,\bm{a}_3$ into $\bm{\tilde a}_1,\bm{\tilde a}_2,\bm{\tilde a}_3$. The two sets are related by integer coefficients $n_{ij}$,
\begin{equation}
\bm{\tilde a}_i=\sum_{j=1}^3 n_{ij}\bm{a}_j,\;\;n_{ij}\in\mathbb{Z}.
\end{equation}
The corresponding primitive vectors $\bm{b}$, $\bm{\tilde b}$ in reciprocal space satisfy
\begin{equation}
\bm{b}_i\cdot\bm{a}_j=2\pi\delta_{ij},\;\;\bm{\tilde b}_i\cdot\bm{\tilde a}_j=2\pi\delta_{ij}.
\end{equation}
Any momentum $\bm{k}$ can thus be expanded as
\begin{equation}
\bm{k}=\frac{1}{2\pi}\sum_{i=1}^3(\bm{\tilde a}_i\cdot\bm{k})\bm{\tilde b}_i=\frac{1}{2\pi}\sum_{i,j=1}^3 n_{ij}(\bm{a}_j\cdot\bm{k})\bm{\tilde b}_i.\label{kexpansion}
\end{equation}

A TRIM $\bm{K}_\alpha$ in the first Brillouin zone of the original lattice is given by
\begin{equation}
\bm{K}_{\alpha}=\tfrac{1}{2}\sum_{i=1}^3 m_{\alpha,i}\bm{b}_i,\;\;m_{\alpha,i}\in\{0,1\}.
\end{equation}
The index $\alpha$ labels each TRIM, identified by the 8 distinct triples $(m_{\alpha,1},m_{\alpha,2},m_{\alpha,3})\in\mathbb{Z}_2\otimes\mathbb{Z}_2\otimes\mathbb{Z}_2$. Subsitution into the expansion \eqref{kexpansion} gives
\begin{align}
\bm{K}_{\alpha}&=\tfrac{1}{2}\sum_{l=1}^3 m_{\alpha,l}\left(\frac{1}{2\pi}\sum_{i,j=1}^3 n_{ij}(\bm{a}_j\cdot\bm{b}_l)\bm{\tilde b}_i\right)\nonumber\\
&=\tfrac{1}{2} \sum_{i,j=1}^3 m_{\alpha,j} n_{ij}\bm{\tilde b}_i.
\end{align}

\begin{table}[htb]
\begin{tabular}{c|cccccccc}
&  \multicolumn{8}{c}{$m_{\alpha,1}\,m_{\alpha,2}\,m_{\alpha,3}$}               \\ 
$n_{i1}\,n_{i2}\,n_{i3}$ (mod 2) &  000 & 001 & 010 & 011 & 100 & 101 & 110 & 111 \\ \hline
  000       &  0   & 0   & 0   & 0   & 0   & 0   & 0   & 0   \\
  001       &  0   & $\tfrac{1}{2}$   & 0   & $\tfrac{1}{2}$   & 0   & $\tfrac{1}{2}$   & 0   & $\tfrac{1}{2}$   \\
 010       &  0   & 0   & $\tfrac{1}{2}$   & $\tfrac{1}{2}$   & 0   & 0   & $\tfrac{1}{2}$   & $\tfrac{1}{2}$   \\
011       &  0   & $\tfrac{1}{2}$   & $\tfrac{1}{2}$   & 0   & 0   & $\tfrac{1}{2}$   & $\tfrac{1}{2}$   & 0   \\
  100       &  0   & 0   & 0   & 0   & $\tfrac{1}{2}$   & $\tfrac{1}{2}$   & $\tfrac{1}{2}$   & $\tfrac{1}{2}$   \\
 101       &  0   & $\tfrac{1}{2}$   & 0   & $\tfrac{1}{2}$   & $\tfrac{1}{2}$   & 0   & $\tfrac{1}{2}$   & 0   \\
  110       &  0   & 0   & $\tfrac{1}{2}$   & $\tfrac{1}{2}$   & $\tfrac{1}{2}$   & $\tfrac{1}{2}$   & 0   & 0   \\
  111       &  0   &  $\tfrac{1}{2}$   & $\tfrac{1}{2}$   & 0   & $\tfrac{1}{2}$   & 0   & 0   & $\tfrac{1}{2}$   
\end{tabular}
\caption{Values of $\nu_{\alpha,i}$ calculated from Eq.\ \eqref{nualpharesult}, for each triple $n_{i1}\,n_{i2}\,n_{i3}$  and each triple $m_{\alpha,1}\,m_{\alpha,2}\,m_{\alpha,3}$ (both $\in\mathbb{Z}_2\otimes\mathbb{Z}_2\otimes\mathbb{Z}_2$).
If we select any two rows and intersect with any column to obtain an ordered pair of values $\nu,\nu'$, we can then find a second column with the same $\nu,\nu'$ at the intersection.
}
\label{table_nu}
\end{table}

We now fold $\bm{K}_\alpha\mapsto \bm{\tilde K}_\alpha$ into the first Brillouin zone of the $\bm{\tilde b}$ reciprocal vectors,
\begin{equation}
\begin{split}
&\bm{\tilde K}_{\alpha}=\sum_{i=1}^3 \nu_{\alpha,i}\bm{\tilde b}_i,\;\;\nu_{\alpha,i}\in[0,1),\\
&\nu_{\alpha,i}=\tfrac{1}{2}\sum_{j=1}^3 m_{\alpha,j}n_{ij}\;(\text{mod}\; 1).
\end{split}\label{nualpharesult}
\end{equation}
In Table \ref{table_nu} we list for each TRIM and each choice of $(n_{i1},n_{i2},n_{i3})\in\mathbb{Z}_2\otimes\mathbb{Z}_2\otimes\mathbb{Z}_2$ the corresponding value of $\nu_{\alpha,i}\in\{0,\tfrac{1}{2}\}$.

We fix the $y$ and $z$-components of $\bm{\tilde K}_{\alpha}$ by specifying $\nu_{\alpha,2}$ and $\nu_{\alpha,3}\in\{0,\tfrac{1}{2}\}$ and ask how many choices of $\alpha$ remain, so how many values of $\alpha$ satisfy the two equations
\begin{equation}
\begin{split}
&\nu_{\alpha,2}=\tfrac{1}{2}\sum_{i=1}^3 n_{2i}m_{\alpha,i}\;(\text{mod}\; 1),\\
&\nu_{\alpha,3}=\tfrac{1}{2}\sum_{i=1}^3 n_{3i}m_{\alpha,i}\;(\text{mod}\; 1).
\end{split}
\end{equation}

Inspection of Table \ref{table_nu} shows that the number of solutions is even. More specifically, there are
\begin{itemize}
\item 8 solutions if
$n_{21},n_{22},n_{23}$ and $n_{31},n_{32},n_{33}$ both equal $000$ mod 2;
\item 4 solutions if only one of $n_{21},n_{22},n_{23}$ and $n_{31},n_{32},n_{33}$ equals $000$ mod 2;
\item 4 solutions if $n_{21},n_{22},n_{23}$ and $n_{31},n_{32},n_{33}$ are identical and different from $000$ mod 2;
\item 2 solutions otherwise. 
\end{itemize}
The multiple solutions correspond to pairs $\bm{K}_\alpha$ and $\bm{K}_\beta$ that are either folded onto the same $\bm{\tilde K}_\alpha=\bm{\tilde K}_\beta$ (if ${\rm det}\,n=0$ mod 2), or onto $\bm{\tilde K}_\alpha$ and $\bm{\tilde K}_\beta$ that differ only in the $x$-component (if ${\rm det}\,n=1$ mod 2). These are the TRIM that are coupled by the boundary normal to the $x$-axis.

\section{Criterion for the appearance of surface Fermi arcs}
\label{app_chirality}

When the boundary couples only Weyl cones of the same chirality, these persist and give rise to surface Fermi arcs. If, however, opposite chiralities are coupled, then the boundary gaps out the Weyl cones and no Fermi arcs appear. Which of these two possibilities is realized can be determined by using that the parity of $m_{\alpha1}+m_{\alpha2}+m_{\alpha3}$ determines the chirality of the Weyl cone at $\bm{K}_\alpha$. 

Table \ref{howmanyopposite} identifies for each choice of $n_{21},n_{22},n_{33}$ and $n_{31},n_{32},n_{33}$ how many pairs of Weyl cones of opposite chirality are folded onto the same point of the surface Brillouin zone. We conclude that surface Fermi arcs appear if either
\begin{itemize}
 \item $n_{2i}+n_{3i}=1$ mod 2 for each $i$, or
 \item $n_{21},n_{22},n_{23}=111$ mod 2, or
 \item $n_{31},n_{32},n_{33}=111$ mod 2.
 \end{itemize}

\begin{table}[htb]
	\begin{tabular}{c|cccccccc}
		&  \multicolumn{8}{c}{	$n_{31}\,n_{32}\,n_{33}$ (mod 2)}  \\ 
		$n_{21}\,n_{22}\,n_{23}$ (mod 2) &  000 & 001 & 010 & 011 & 100 & 101 & 110 & 111 \\ \hline
		000       & 4   & 2   & 2   & 2   & 2   & 2   & 2   & 0   \\
		001       & 2   & 2   & 1   & 1   & 1   & 1   & 0   & 0   \\
		010       & 2   & 1   & 2   & 1   & 1   & 0   & 1   & 0   \\
		011       & 2   & 1   & 1   & 2   & 0   & 1   & 1   & 0   \\
		100       & 2   & 1   & 1   & 0   & 2   & 1   & 1   & 0   \\
		101       & 2   & 1   & 0   & 1   & 1   & 2   & 1   & 0   \\
		110       & 2   & 0   & 1   & 1   & 1   & 1   & 2   & 0   \\
		111       & 0   & 0   & 0   & 0   & 0   & 0   & 0   & 0   
	\end{tabular}
	\caption{\label{howmanyopposite} Number of pairs of opposite-chirality Weyl cones that are coupled by a surface termination characterized by the integers $n_{2i},n_{3i}$, $i\in\{1,2,3\}$. When this number equals 0 the surface couples only Weyl cones of the same chirality and surface Fermi arcs will appear. If the number is different from zero the surface does not support Fermi arcs.}
\end{table}

\section{Calculation of the dispersive Landau bands due to the coupling of open and closed orbits}
\label{sec_spectrum}

\begin{figure}[b]
\centerline{\includegraphics[width=0.9\linewidth]{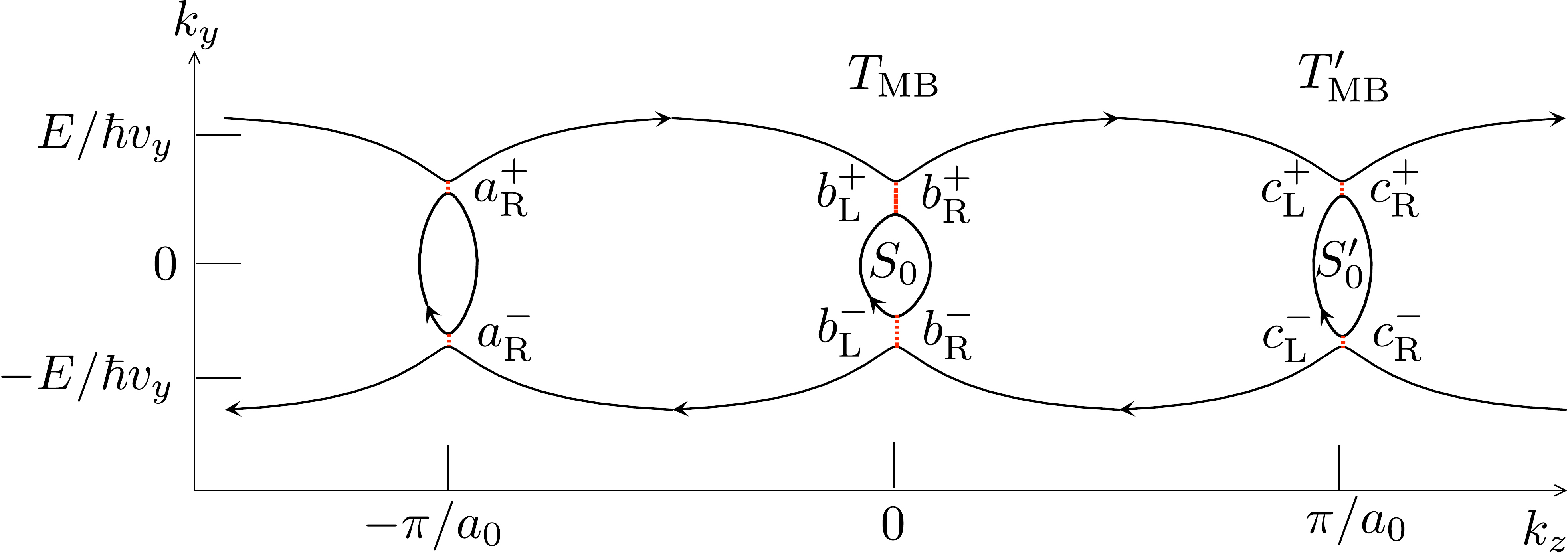}}
\caption{Equi-energy contours in the $k_y$--$k_z$ plane. The labeled wave amplitudes are related by the scattering and transfer matrices \eqref{scattering_matrix}--\eqref{Ttotalresult}.
}
\label{fig_chain}
\end{figure}

\begin{figure*}[tb]
\centerline{\includegraphics[width=1\linewidth]{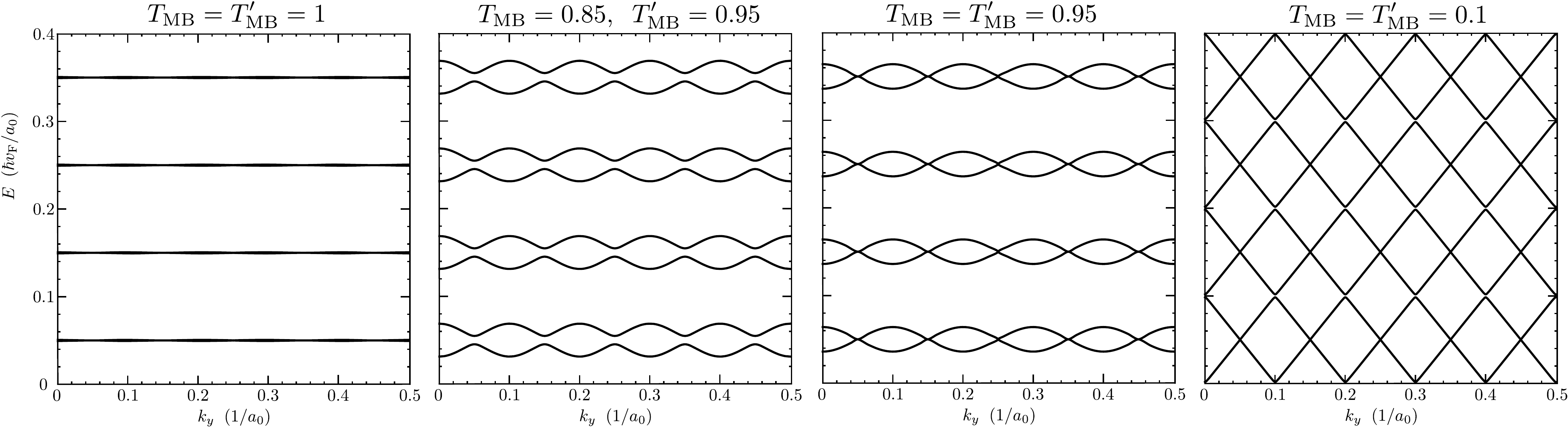}}
\caption{Dispersion relation of the slab in a perpendicular magnetic field, calculated from Eqs.\ \eqref{Ttotalresulttr} and \eqref{trTtotalcosqL} for $W=1.8\,a_0$, $S_0$=$S'_0$, $\nu=1/2$, $B=0.1\,\hbar/ea_0^2$. The four panels correspond to different choices of the magnetic breakdown probabilities $T_{\rm MB}$ and $T'_{\rm MB}$ at the two Weyl points. At the two extremes of strong and weak magnetic breakdown we see dispersionless Landau levels (left-most panel) and linearly dispersing surface modes (right-most panel).
}
\label{fig_fourplots}
\end{figure*}

To calculate the effect of the coupling of open and closed orbits on the Landau levels we apply the scattering theory of Refs.\ \onlinecite{Kag83,Gvo07,Gvo86} to the equi-energy contours shown in Fig.\ \ref{fig_chain}. We distinguish the two Weyl points at $k_z=0$ and $k_z=\pi/a_0$ by their different magnetic breakdown probability, denoted respectively by $T_{\rm MB}=1-R_{\rm MB}$ and $T'_{\rm MB}=1-R'_{\rm MB}$. The areas of the closed orbits may also differ, we denote these by $S_0$ and $S'_0$ and the corresponding phase shifts by $\phi=S_0l_m^2+2\pi\nu$ and $\phi'=S'_0l_m^2+2\pi\nu$. 

The coupling of the closed and open orbits at these two Weyl points is described by a pair of scattering matrices, given by
\begin{subequations}
\label{scattering_matrix}
\begin{align}
&\begin{pmatrix}
b_{\rm L}^-\\
b_{\rm R}^+
\end{pmatrix}=\begin{pmatrix}
r&t\\
t&r
\end{pmatrix}\cdot\begin{pmatrix}
b_{\rm L}^+\\
b_{\rm R}^-
\end{pmatrix},\;\;
r=\frac{T_{\rm MB}e^{i\phi/2}}{1-R_{\rm MB}e^{i\phi}},\\
&t=-\sqrt{R}_{\rm MB}+\frac{T_{\rm MB}\sqrt{R}_{\rm MB}e^{i\phi}}{1-R_{\rm MB}e^{i\phi}},
\end{align}
\end{subequations}
for the Weyl point at $k_z=0$, and similarly for the other Weyl point at $k_z=\pi/a_0$ (with $T_{\rm MB}\mapsto T'_{\rm MB}$, $\phi\mapsto\phi'$). The coefficients can be rearranged in an energy-dependent transfer matrix,
\begin{equation}
\begin{pmatrix}
b_{\rm R}^+\\
b_{\rm R}^-
\end{pmatrix}={\cal T}(E)\begin{pmatrix}
b_{\rm L}^+\\
b_{\rm L}^-
\end{pmatrix},\;\;
{\cal T}=\begin{pmatrix}
t-r^2/t&r/t\\
-r/t&1/t
\end{pmatrix},
\end{equation}
and similarly for ${\cal T}'$ (with $t\mapsto t'$, $r\mapsto r'$). The transfer matrices are energy dependent via the energy dependence of $S_0$ and hence of $\phi$.

We ignore the curvature of the open orbits, approximating them by straight contours along the line $k_y=E/\hbar v_{y}$. The phase shift accumulated upon propagation from one Weyl point to the next, in the Landau gauge $\bm{A}=(0,-Bz,0)$, is then given by 
\begin{equation}
\psi= \frac{E}{\hbar v_{y}}\frac{\pi}{a_0}l_m^2= \frac{\pi E}{\hbar\omega_c},\;\; \omega_c=eBv_y a_0/\hbar.
\end{equation}
The full transfer matrix over the first Brillouin zone takes the form
\begin{widetext}
\begin{align}
\begin{pmatrix}
c_{\rm R}^+\\
c_{\rm R}^-
\end{pmatrix}={}&{\cal T}_{\rm total}(E)\begin{pmatrix}
a_{\rm R}^+\\
a_{\rm R}^-
\end{pmatrix},\;\;
{\cal T}_{\rm total}=\begin{pmatrix}
t'-r'^2/t'&r'/t'\\
-r'/t'&1/t'
\end{pmatrix}\begin{pmatrix}
e^{i\psi}&0\\
0&e^{-i\psi} \end{pmatrix}
\begin{pmatrix}
t-r^2/t&r/t\\
-r/t&1/t
\end{pmatrix}\begin{pmatrix}
e^{i\psi}&0\\
0&e^{-i\psi}
\end{pmatrix},\label{Ttotalresult}
\\
{\rm tr}\,{\cal T}_{\rm total}={}&\frac{ \bigl(e^{i {\phi}}-R_{\rm MB}\bigr) \bigl(e^{i {\phi'}}-R'_{\rm MB}\bigr)+\bigl(1-e^{i {\phi}} R_{\rm MB}\bigr) \bigl(1-e^{i {\phi'}} R'_{\rm MB}\bigr)-2 T_{\rm MB} T'_{\rm MB} e^{\frac{1}{2} i ({\phi}+{\phi'})+2i\psi}}{e^{2i\psi}\bigl(e^{i {\phi}}-1\bigr) \bigl(e^{i {\phi'}}-1\bigr) \sqrt{R_{\rm MB}R'_{\rm MB}}}.\label{Ttotalresulttr}
\end{align}
\end{widetext}

Because ${\rm det}\,{\cal T}_{\rm total}=1$, the eigenvalues of ${\cal T}_{\rm total}$ come in inverse pairs $\lambda,1/\lambda$. The transfer matrix translates the wave function over a period ${\cal L}$ in real space, so we require that $\lambda=e^{iq{\cal L}}$ for some real wave number $q$, hence $\lambda+1/\lambda=e^{iq{\cal L}}+e^{-iq{\cal L}}$, or equivalently \cite{Gvo86}
\begin{equation}
{\rm tr}\,{\cal T}_{\rm total}(E)=2\cos q{\cal L}.\label{trTtotalcosqL}
\end{equation}
(In the main text we denote $q$ by $k_y$, here we choose a different symbol as a reminder that $q$ is a conserved quantity, while the zero-field wave vector is not.) A numerical solution of Eq.\ \eqref{trTtotalcosqL} is shown in Figs.\ \ref{fig_oscillation} and \ref{fig_fourplots}.

For $T_{\rm MB}$ and $T'_{\rm MB}$ close to unity an analytical solution $E_n(q)$ for the dispersive Landau bands can be obtained. We substitute $\psi=\pi(n-\nu)-(\phi+\phi')/4+\pi\delta E/\hbar\omega_c$ into Eq.\ \eqref{Ttotalresulttr} and expand to second order in $\delta E$ and to first order in $R_{\rm MB},R'_{\rm MB}$. Then we equate to $2\cos q{\cal L}$ to arrive at
\begin{subequations}
\label{Enqsimple}
\begin{align}
&E^\pm_n(q)=(n-\nu)\hbar\omega_c\pm \delta E(q),\\
&(\pi\delta E/\hbar \omega_c)^2=\rho+\rho'+2\sqrt{\rho\rho'}\cos q{\cal L},\\
&\rho=R_{\rm MB}\sin^2(\phi/2),\;\;\rho'=R'_{\rm MB}\sin^2(\phi'/2),
\label{Enqsimpleb}
\end{align}
\end{subequations}
where $\phi$ and $\phi'$ are evaluated at $E=(n-\nu)\hbar\omega_c$. Corrections are of second order in $R_{\rm MB}$ and $R'_{\rm MB}$ and we have assumed that the areas $S_0,S'_0$ of the closed orbit are small compared to $k_{\rm F}/a_0$ --- so that variations of $\phi$ and $\phi'$ over the Landau band can be neglected relative to the band spacing $\hbar\omega_c$.

\section{Landau levels from surface Fermi arcs}
\label{sec_landauoffset}

\begin{figure}[tb]
	\includegraphics[width=0.6\linewidth]{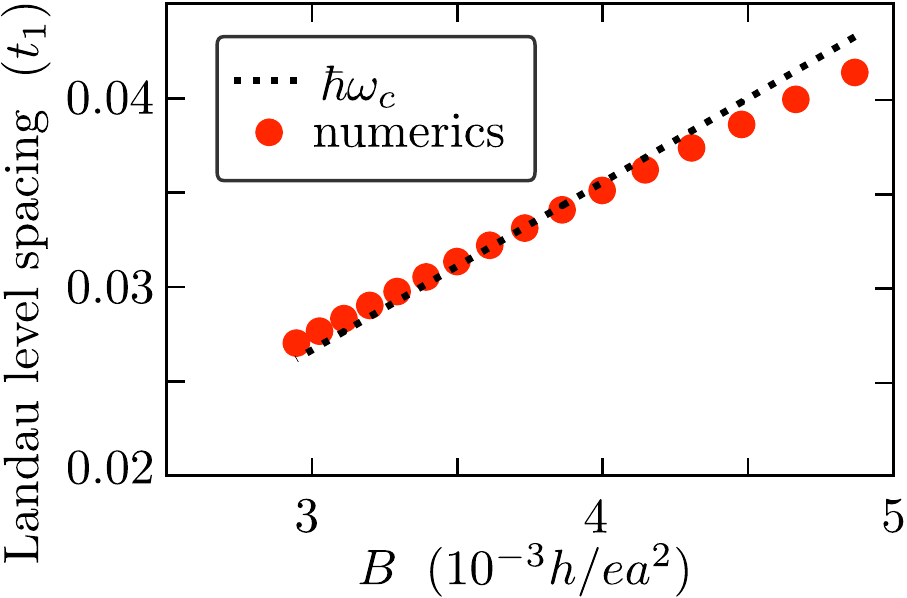}
	\caption{Magnetic field dependence of the energy spacing of the Landau levels near $E=0$. The numerical data is for the slab geometry of Fig.\ \ref{fig_tb_dispLL} ($W=11a'$, $L=30a'$) at $t_0=\delta t_0= 0$ so that the probability of magnetic breakdown is unity and the Landau levels are dispersionless. The predicted energy spacing $\hbar\omega_c=eBv_{\rm F}a'$ is the black dotted line.}
	\label{fig_spacinglandau}
\end{figure}

\begin{figure}[tb]
	\includegraphics[width=0.8\linewidth]{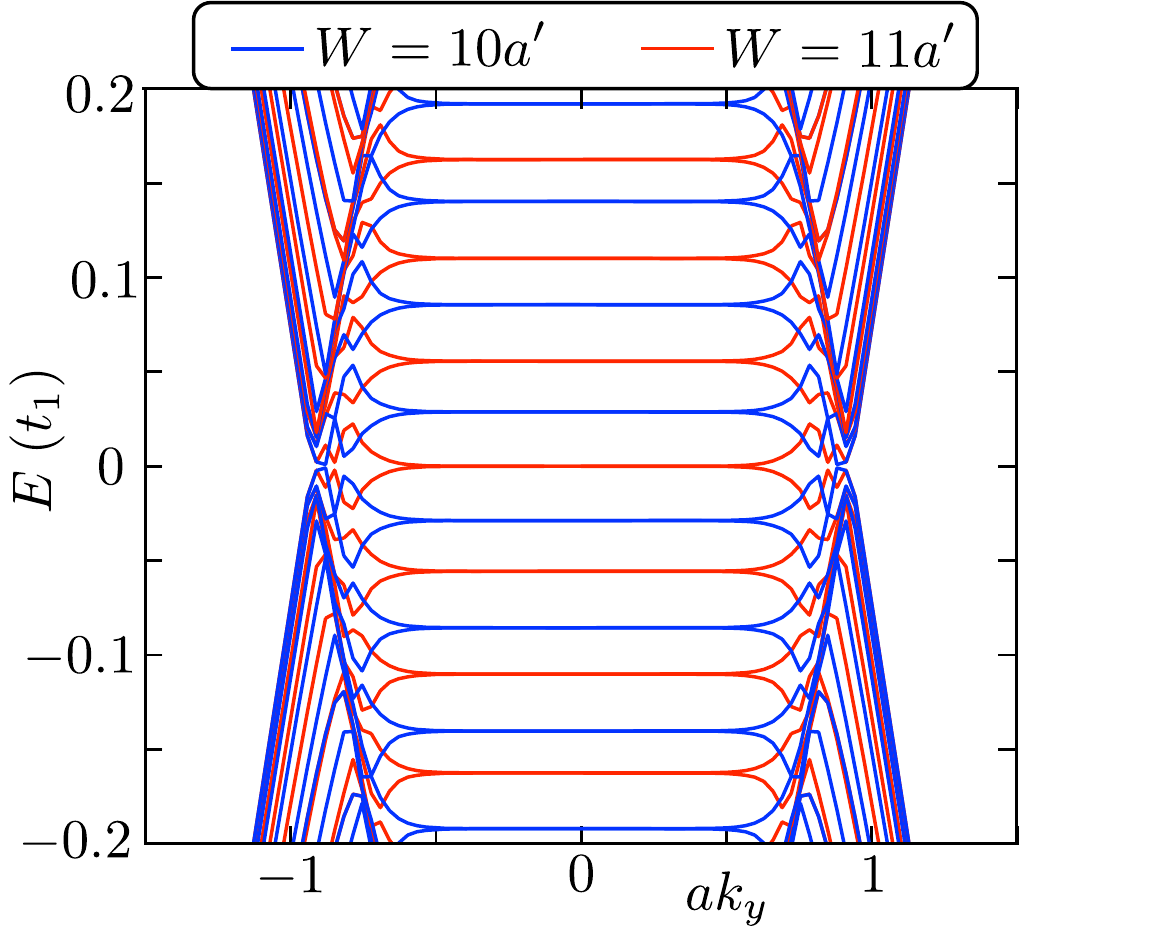}
	\caption{Dispersion relation of the tight-binding model with $t_0=\delta t_0= 0$, for $B=7.07\cdot 10^{-3}\,h/ea^2$, $L=30a'$, and two values of $W=10a'$ and $11a'$. The Landau levels are shifted by half a level spacing when $W/a'$ switches from odd to even, indicating a shift of the offset $\nu$ from 0 to $1/2$.}
	\label{fig_offsetlandau}
\end{figure}

As explained in Fig.\ \ref{fig_contours}, the spacing of Landau levels formed out of surface Fermi arcs varies $\propto B$ --- in contrast to the $\sqrt{B}$ dependence for unconfined massless electrons. In the tight-binding model of Sec.\ \ref{sec_numerics} we can test this by setting $\varepsilon=\varepsilon'=0$, so that there are only closed orbits and the Landau levels are dispersionless. The expected quantization is
\begin{equation}
E_n=(n-\nu)\hbar\omega_c,\;\;\omega_c=eBv_{\rm F} a'/\hbar,\;\;n=0,1,2,\ldots
\end{equation}
with $v_{\rm F}$ the velocity in the surface Fermi arc, connecting Weyl points spaced by $\pi/a'$. As shown in Fig.\ \ref{fig_spacinglandau}, this agrees nicely with the numerics.

In an unconfined 2D electron gas, the offset $\nu$ equals 1/2 or 0 for massive or massless electrons, respectively. For the surface Fermi arcs we observe that $\nu$ depends on the parity of the number of unit cells between top and bottom surface: $\nu=0$ if $W/a'$ is odd, while $\nu=1/2$ if $W/a'$ is even. This parity effect suggests that the coupling of Fermi arc states on opposite surfaces, needed to close the orbit in Fig.\ \ref{fig_conveyor}, introduces a phase shift that depends on the parity of $W/a'$. We are not aware of such a phase shift for generic Weyl semimetals \cite{Pot14,Zha16,Ale17,Bre18}, it seems to be a characteristic feature of Kramers-Weyl fermions that deserves further study.


\clearpage

\begin{thebibliography}{99}
\bibitem{Cha18} G. Chang, B. J. Wieder, F. Schindler, D. S. Sanchez, I. Belopolski, S.-M. Huang, B. Singh, D. Wu, T. Neupert, T.-R. Chang, S.-Y. Xu, H. Lin, and M. Z. Hasan, \textit{Topological quantum properties of chiral crystals}, Nature Mat. \textbf{17}, 978 (2018).
\bibitem{She18} C. Shekhar, \textit{Chirality meets topology}, Nature Mat. \textbf{17}, 953 (2018).
\bibitem{Rao19} Zhicheng Rao, Hang Li, Tiantian Zhang, Shangjie Tian, Chenghe Li, Binbin Fu, Cenyao Tang, Le Wang, Zhilin Li, Wenhui Fan, Jiajun Li, Yaobo Huang, Zhehong Liu, Youwen Long, Chen Fang, Hongming Weng, Youguo Shi, Hechang Lei, Yujie Sun, Tian Qian, and Hong Ding, \textit{Observation of unconventional chiral fermions with long Fermi arcs in CoSi}, Nature \textbf{567}, 496 (2019).
\bibitem{San19} D. S. Sanchez, I. Belopolski, T. A. Cochran, X. Xu, J.-X. Yin, G. Chang, W. Xie, K. Manna, V. S\"{u}{\ss}, C.-Y. Huang, N. Alidoust, D. Multer, S. S. Zhang, N. Shumiya, X. Wang, G.-Q. Wang, T.-R. Chang, C. Felser, S.-Y. Xu, S. Jia, H. Lin, M. Z. Hasan, \textit{Topological chiral crystals with helicoid-arc quantum states}, Nature \textbf{567}, 500 (2019).
\bibitem{Tak19} D. Takane, Z. Wang, S. Souma, K. Nakayama, T. Nakamura, H. Oinuma, Y. Nakata, H. Iwasawa, C. Cacho, T. Kim, K. Horiba, H. Kumigashira, T. Takahashi, Y. Ando, T. Sato, \textit{Observation of chiral fermions with a large topological charge and associated Fermi-arc surface states in CoSi}, Phys. Rev. Lett. \textbf{122}, 076402 (2019).
\bibitem{Yua19} Qian-Qian Yuan, Liqin Zhou, Zhi-Cheng Rao, Shangjie Tian, Wei-Min Zhao, Cheng-Long Xue, Yixuan Liu, Tiantian Zhang, Cen-Yao Tang, Zhi-Qiang Shi, Zhen-Yu Jia1, Hongming Weng, Hong Ding, Yu-Jie Sun, Hechang Lei, and Shao-Chun Li, \textit{Quasiparticle interference evidence of the topological Fermi arc states in chiral fermionic semimetal CoSi}, Science Adv. \textbf{5}, eaaw9485 (2019).
\bibitem{Sch19} N. B. M. Schr\"{o}ter, D. Pei, M. G. Vergniory, Y. Sun, K. Manna, F. de Juan, J.. A. Krieger, V. S\"{u}ss, M. Schmidt, P. Dudin, B. Bradlyn, T. K. Kim, T. Schmitt, C. Cacho, C. Felser, V. N. Strocov, and Y. Chen, \textit{Chiral topological semimetal with multifold band crossings and long Fermi arcs}, Nature Phys. \textbf{15}, 759 (2019).
\bibitem{Zha17} C.-L. Zhang, F. Schindler, H. Liu, T.-R. Chang, S.-Y. Xu, G. Chang, W. Hua, H. Jiang, Z. Yuan, J. Sun, H.-T. Jeng, H.-Z. Lu, H. Lin, M. Z. Hasan, X. C. Xie, T. Neupert, and S. Jia, \textit{Ultraquantum magnetoresistance in Kramers Weyl semimetal candidate $\beta$--Ag$_2$Se}, Phys. Rev. B \textbf{96}, 165148 (2017).
\bibitem{Wa18} B. Wan, F. Schindler, K. Wang, K. Wu, X. Wan, T. Neupert, and H.-Z. Lu, \textit{Theory for the negative longitudinal magnetoresistance in the quantum limit of Kramers Weyl semimetals}, J. Phys. Condens. Matter \textbf{30}, 505501 (2018).
\bibitem{He19} Wen-Yu He, Xiao Yan Xu, K. T. Law, \textit{Kramers Weyl semimetals as quantum solenoids and their applications in spin-orbit torque devices}, arXiv:1905.12575.
\bibitem{Ger89} R. R. Gerhardts, D. Weiss, and K. von Klitzing, \textit{Novel magnetoresistance oscillations in a periodically modulated two-dimensional electron gas}, Phys. Rev. Lett. \textbf{62}, 1173 (1989).
\bibitem{Win89} R. W. Winkler, J. P, Kotthaus, and K. Ploog, \textit{Landau band conductivity in a two-dimensional electron system modulated by an artificial one-dimensional superlattice potential}, Phys. Rev. Lett. \textbf{62}, 1177 (1989).
\bibitem{Bee89} C. W. J. Beenakker, \textit{Guiding-center-drift resonance in a periodically modulated two-dimensional electron gas}, Phys. Rev. Lett. \textbf{62}, 2020 (1989).
\bibitem{Str90} P. St\v{r}eda and A. H. MacDonald, \textit{Magnetic breakdown and magnetoresistance oscillations in a periodically modulated two-dimensional electron gas}, Phys. Rev. B \textbf{41}, 11892 (1990).
\bibitem{Gvo07} V. M. Gvozdikov, \textit{Magnetoresistance oscillations in a periodically modulated two-dimensional electron gas: The magnetic-breakdown approach}, Phys. Rev. B \textbf{75}, 115106 (2007).
\bibitem{Pot14} A. C. Potter, I. Kimchi, and A. Vishwanath, \textit{Quantum oscillations from surface Fermi-arcs in Weyl and Dirac semi-metals}, Nature Comm. \textbf{5}, 5161 (2014).
\bibitem{Zha16} Y. Zhang, D. Bulmash, P. Hosur, A. C. Potter, and A. Vishwanath, \textit{Quantum oscillations from generic surface Fermi arcs and bulk chiral modes in Weyl semimetals}, Sci. Rep. \textbf{6}, 23741 (2016).
\bibitem{Akh08} A. R. Akhmerov and C. W. J. Beenakker, \textit{Boundary conditions for Dirac fermions on a terminated honeycomb lattice}, Phys. Rev. B \textbf{77}, 085423 (2008).
\bibitem{Has17} M. Z. Hasan, S.-Y. Xu, I. Belopolski, and S.-M. Huang, \textit{Discovery of Weyl fermion semimetals and topological Fermi arc states}, Annu. Rev. Condens. Matter Phys. \textbf{8}, 289(2017).
\bibitem{Yan17} B. Yan and C. Felser, \textit{Topological Materials: Weyl Semimetals}, Annu. Rev. Condens. Matter Phys. \textbf{8}, 337 (2017).
\bibitem{Bur18} A. A. Burkov, \textit{Weyl Metals}, Annu. Rev. Condens. Matter Phys. \textbf{9}, 359 (2018).
\bibitem{Arm18} N. P. Armitage, E. J. Mele, and A. Vishwanath, \textit{Weyl and Dirac semimetals in three-dimensional solids}, Rev. Mod. Phys. \textbf{90}, 15001 (2018).
\bibitem{Bov18} N. Bovenzi, M. Breitkreiz, T. E. O'Brien, J. Tworzyd{\l}o, and C. W. J. Beenakker, \textit{Twisted Fermi surface of a thin-film Weyl semimetal}, New J. Phys. \textbf{20}, 023023 (2018).
\bibitem{Bar17} V. Barsan and V. Kuncser, \textit{Exact and approximate analytical solutions of Weiss equation of ferromagnetism and their experimental relevance}, Phil. Mag. Lett. \textbf{97}, 359 (2017).
\bibitem{Pip69} A. B. Pippard, \textit{Magnetic breakdown}, in: \textit{Physics of Solids in Intense Magnetic Fields} (Springer, Boston, 1969).
\bibitem{Kag83} M. I. Kaganov and A. A. Slutskin, \textit{Coherent magnetic breakdown}, Phys. Rep. \textbf{98}, 189 (1983).
\bibitem{Sta67} R. W. Stark and L. M. Falicov, \textit{Magnetic breakdown in metals}, Prog. Low Temp. Phys \textbf{5}, 235 (1967).
\bibitem{Gvo86} V. M. Gvozdikov, \textit{Thermodynamic oscillations in periodic magnetic breakdown structures}, Fiz. Nizk. Temp. \textbf{12}, 705 (1986).
\bibitem{kwant} C. W. Groth, M. Wimmer, A. R. Akhmerov, and X. Waintal, \textit{Kwant: A software package for quantum transport}, New J. Phys. \textbf{16}, 063065 (2014).
\bibitem{Ale17} A. Alexandradinata and L. Glazman, \textit{Geometric phase and orbital moment in quantization rules for magnetic breakdown}, Phys. Rev. Lett. \textbf{119}, 256601 (2017).
\bibitem{Bre18} M. Breitkreiz, N. Bovenzi, and J. Tworzyd{\l}o, \textit{Phase shift of cyclotron orbits at type-I and type-II multi-Weyl nodes}, Phys. Rev. B \textbf{98}, 121403 (2018).

\end{thebibliography}
\end{document}